\documentclass[final,leqno]{siamltex}

\usepackage{etex,latexsym,epsfig,amssymb,amsmath,subfigure,color,pictex,amsmath,listings,algorithmic,algorithm,tabularx}

\baselineskip{18pt}

\renewcommand{\d}{\mathrm{d}}

\title{A mathematical model for mechanically-induced deterioration of the binder in lithium-ion electrodes}
\author{
J. M. Foster$^{\mbox{\S,}}$\thanks{Department of Mathematics \& Statistics, McMaster University, Hamilton, Canada, L8S 4K1 (jamie.foster@port.ac.uk, bprotas@mcmaster.ca). $^{\mbox{\S}}$Current address: Department of Mathematics, University of Portsmouth, Portsmouth, UK, PO1 2UP.}
\and S. J. Chapman\thanks{Mathematical Institute, University of Oxford, Oxford, UK, OX2 6GG.}
\and G. Richardson\thanks{School of Mathematics, University of Southampton, Southampton, UK, SO17 1BJ.}
\and B. Protas$^*$.}

\usepackage[english]{babel}

\newcommand{\bes} {\begin{eqnarray*}}
\newcommand{\ees} {\end{eqnarray*}}

\newcommand{\be}{\begin{eqnarray}}
\newcommand{\ee}{\end{eqnarray}}

\newcommand{\dd} \partial

\newcommand{\ie}{{\it i.e.}\ }

\newcommand{\etal}{{\it et~al.}\ }
\newcommand{\ra}{\rightarrow }

\begin{document}
\maketitle

\section*{Abstract}

This study is concerned with modeling detrimental deformations of the binder phase within lithium-ion batteries that occur during cell assembly and usage. A two-dimensional poroviscoelastic model for the mechanical behavior of porous electrodes is formulated and posed on a geometry corresponding to a thin rectangular electrode, with a regular square array of microscopic circular electrode particles, stuck to a rigid base formed by the current collector. Deformation is forced both by (i) electrolyte absorption driven binder swelling, and; (ii) cyclic growth and shrinkage of electrode particles as the battery is charged and discharged. In order to deal with the complexity of the geometry the governing equations are upscaled to obtain macroscopic effective-medium equations. A solution to these equations is obtained, in the asymptotic limit that the height of the rectangular electrode is much smaller than its width, that shows the macroscopic deformation is one-dimensional, with growth confined to the vertical direction. The confinement of macroscopic deformations to one dimension is used to obtain boundary conditions on the microscopic problem for the deformations in a 'unit cell' centered on a single electrode particle. The resulting microscale problem is solved using numerical (finite element) techniques. The two different forcing mechanisms are found to cause distinctly different patterns of deformation within the microstructure. Swelling of the binder induces stresses that tend to lead to binder delamination from the electrode particle surfaces in a direction parallel to the current collector, whilst cycling causes stresses that tend to lead to delamination orthogonal to that caused by swelling. The differences between the cycling-induced damage in both: (i) anodes and cathodes, and; (ii) fast and slow cycling are discussed. Finally, the model predictions are compared to microscopy images of nickel manganese cobalt oxide cathodes and a qualitative agreement is found. 
\section{\label{intro}Introduction}

Much current research is focused on the development of lithium-ion
batteries for use in a variety of areas, ranging from consumer
electronics to the automotive industry, with the current market for
such devices being in excess of \$20bn (USD) per annum and set to increase
further \cite{jom}. Many previous studies have considered how to model the
electrochemical processes occurring within such cells, with the aim of
informing optimal design. The majority of these studies have been, at
least partially, based on the seminal models of Newman \etal
\cite{newmana,newmanb,newmanc,newmand}, which account for the complex
battery microstructure using `averaged' quantities---an approach that
has been subsequently formalized in \cite{rdp,ciucci}. Providing an
exhaustive list, or even a summary, of the extant work on lithium-ion
batteries is almost impossible and instead we point to the following
reviews \cite{newman_book,otherbook,bardbook}. Of particular interest
is modeling that can inform design improvements in terms of increased
energy density, facilitation of higher charge/discharge rates, and
improved safety and cycling lifetime. Here, we will focus on the last
of these, specifically the (detrimental) morphological changes that
can occur during cell fabrication and usage as a result of the
development of internal mechanical stresses in response to (i) polymer
binder swelling due to electrolyte absorption, and; (ii) dilation (and
contraction) of the electrode particles as the cell is cycled.

\begin{figure}          \centering
\includegraphics[width=0.7\textwidth]{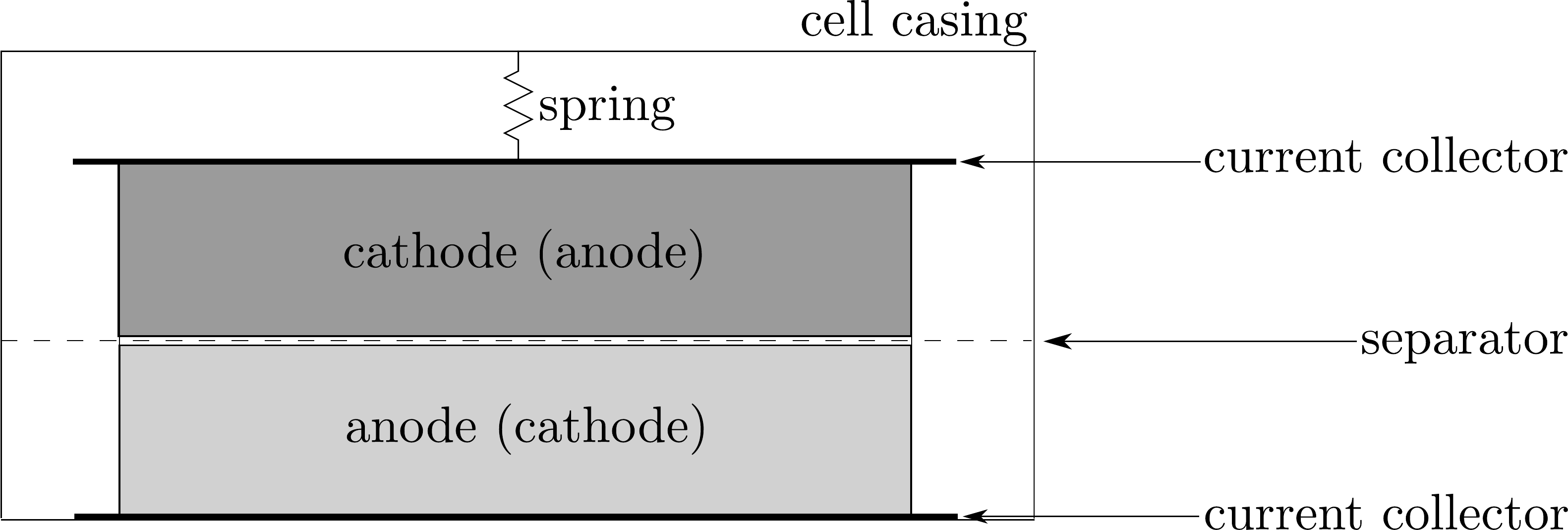}
\label{coin_cell}
\caption{A schematic of a coin cell.}
\end{figure}

\begin{figure}          \centering
\includegraphics[width=0.8\textwidth]{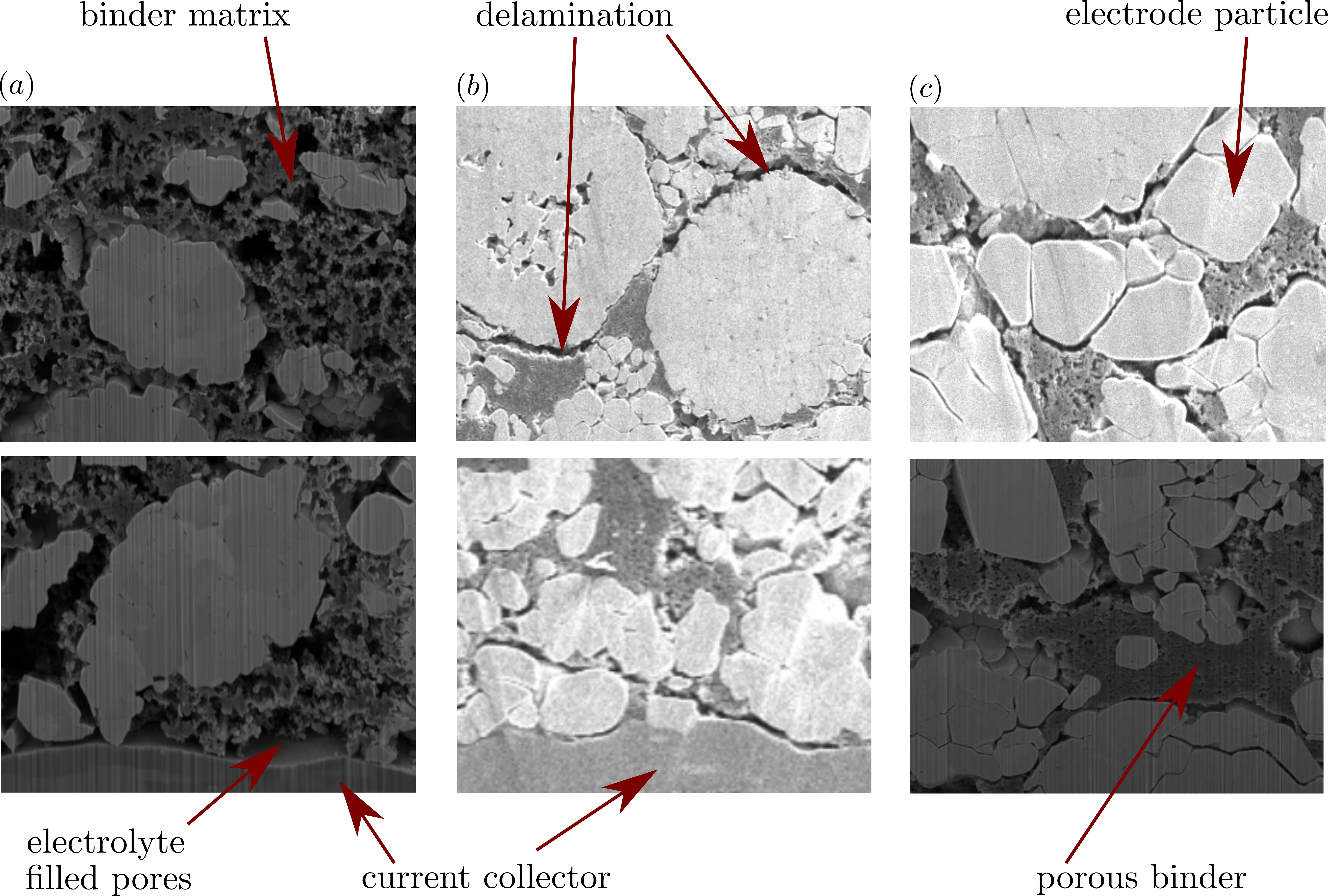}
\caption{FIB/SEM images of nickel manganese cobalt oxide (NMC)
  cathodes with a PVDF binder. The lightest material is the NMC active
  material, the darker material is the polymer binder, the darkest
  regions are pore spaces which are filled with electrolyte during
  cell operation, and the (almost) planar material at the bottom of
  columns (a) and (b) is the aluminum CC. Column (a) shows electrodes
  that were manufactured in G. Goward's group (McMaster Chemistry).
  These images were taken before the electrodes had been
  immersed in electrolyte, i.e.~before they were constructed as part
  of a cell.  Column (b) shows commercial electrodes which have been
  immersed in electrolyte (1 molar LiPF$_6$ in EC/DMC), undergone a
  single `formation' cycle, and then been removed from their housing
  and allowed to dry. Column (c) also shows commercial electrodes, but
  these have been immersed in electrolyte (again, 1 molar LiPF$_6$ in
  EC/DMC), have undergone between 20 and 50 complete charge/discharge
  cycles, and have then been removed from their housing and allowed to
  dry. More example images of these same electrodes can be found in
  \cite{nester,hanshuo}. The images were harvested by G. Botton's
  group, McMaster University, using the procedures described in \cite{hanshuo}. The regions 
  shown in each panel are \emph{approximately}
  10$\mu$m$\times$10$\mu$m. \label{example_slices}}
\end{figure}

Lithium-ion batteries comprise four key constituents: (i) the anode
(the negative electrode during discharge); (ii) the cathode (the
positive electrode during discharge); (iii) the electrolyte, and; (iv)
a porous separator that ensures unhindered passage of ionic current
(via the electrolyte) while electronically isolating the two
electrodes, see Figure \ref{coin_cell}.1. This assembly is sandwiched
between two metallic current collectors (CCs). In commercial cells
both the anode and cathode have a complex morphology; both are
typically composed of small particles ($\sim$1-10$\mu$m) of active
material (AM) embedded in a porous polymer binder matrix which (in the
case of the cathode) is doped with highly conducting acetylene black
nano-particles. The pores within the polymer matrix (of typical size
10--100nm) form channels through which the electrolyte can penetrate
--- see Figure \ref{example_slices}. The binder performs two key
tasks; firstly it ensures the structural integrity of the electrode
and secondly it serves to connect electrode particles electronically
to the current collector.

After manufacture, when such electrodes are assembled as part of a
cell, they are soaked in electrolytic fluid which is absorbed by the
binder and results in significant swelling of this polymer phase.
Commonly used binders include polyvinylidene fluoride (PVDF) and
carboxymethyl cellulose (CMC), and each of these materials is
known to exhibit expansions of around $50\%$ by volume ($75\%$ by
weight) after begin immersed in EC:DMC for a few hours/days
\cite{chen,yash,park}.  Even though this binder/acetylene black phase only
constitutes a relatively small fraction of the material making up an
electrode --- typical figures for the AM:filler ratio by weight (here
filler denotes both polymer binder and acetylene black) range from
70:30 to 90:10 in cathodes \cite{li,liu} and 90:10 to 95:5 in anodes
\cite{buqa,yoo} --- a 50\% volumetric expansion has the potential to
lead to significant deformations and stresses (and ultimately
morphological damage to the electrode) \cite{somethingnew}.

After assembly, when the cell is under operating conditions, the
chemical potential difference between the positive and the negative
electrodes leads to charge transfer events (redox/(de-)intercalation
reactions) in the electrode particles. Importantly, these reactions
are accompanied by volumetric changes in the electrode particles. By
far the most commonly used active material in commercial anodes is
graphitic carbon (LiC$_6$). On lithiation (associated with battery
charging) these anode particles exhibit a significant volumetric
expansion, of around 10\% \cite{dahn,reyn}. In contrast typical
cathode active materials ({e.g.} cobalt oxide (LiCoO$_2$), iron
phosphate (LiFePO$_4$) and nickel manganese cobalt oxide
(LiNiMnCoO$_2$)) exhibit smaller, but nonetheless appreciable,
volumetric expansions in the range of 2--4\% \cite{yabu,ohzu}.
This swelling/contraction of the active material phase can lead to
large mechanical stresses within the electrodes which, in turn, can
lead to delamination of the binder from the electrode particles and
the CC.  Notably delamination, as seen in Figure \ref{example_slices}
occurs preferentially (at least early on) in the plane parallel to the
current collector and it is one of the goals of this work to explain
this pattern of delamination.

In unfavorable cases, structural change within the electrode has
detrimental effects on cell performance. For example, delamination of
the binder from the surfaces of the electrode particles and CC (as can
be seen in Figure \ref{example_slices}) has been proposed as a cause
of degradation, and has been shown to cause both a measurable decrease
in the `connectivity' and the bulk (or `effective') electronic
conductivity of cathodes \cite{nester,scott2,gul14,kerl,hanshuo,scott1,vett}. In
addition to decreasing the efficacy of electron conduction, the
delamination process can increase current densities (by reducing the
area of electronic contact) and in turn enhance Joule heating, an
effect which has the potential to cause to thermal runaway
\cite{wang}. Delamination also exposes additional areas of the
surfaces of the electrode particles directly to the electrolyte
leading to the formation of extra solid electrolyte interface (SEI)
layers \cite{pins} and to the consumption of active lithium (from the
electrolyte), an effect that results in decreased battery capacity.

\begin{figure}          \centering
\includegraphics[width=0.75\textwidth]{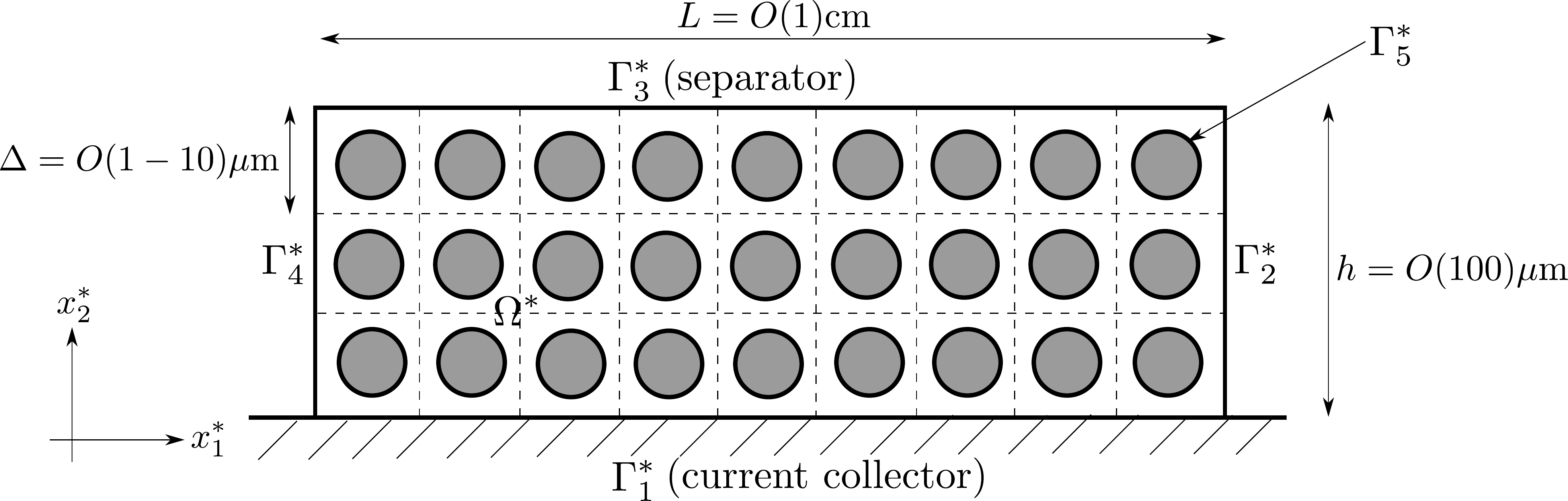}
\caption{A sketch of the idealized model electrode geometry. Explicit definitions of the different boundary segments are given in \S\ref{geometry}. }
\label{sketch}
\end{figure}

Whilst a good deal of work has been focused on: (i) characterizing the
mechanical processes occurring \emph{within} various different types
of active material, {e.g.} phase changes and particle fracturing
\cite{qi,chak1,chak2,baza1,woodford}; (ii) discerning the impact of cycling-induced compression of, and the consequent reduction of pore space in, separator membranes on long-term cell performance \cite{arnold1,arnold2,arnold3}, and; (iii) understanding the relatinship between the stack pressure, in e.g. commerical pouch cells, and the health of the cells \cite{arnold4,arnold5,arnold6}; much less has been done to elucidate the
mechanical behavior of individual electrodes (including the binder) as a whole.
Previous modeling in this area includes \cite{ott,xiaosong2}. The
former considers the effects of electrode particle expansion and
contraction on electrical connectivity when the surrounding matrix is
composed entirely of carbon black particles. However, the absence of a
binder material means these results are irrelevant to commercial
cells. In contrast, whilst \cite{xiaosong2} does model the binder
material, it focuses almost entirely on predicting stresses within the
electrode particles. Here we focus primarily on modeling the 
deformations of the binder itself in response to: (i) its
swelling in response to electrolyte absorption, and; (ii)
swelling/contraction of the electrode particles as the cell is cycled.
We work within an idealized 2-dimensional geometry as illustrated in
Figure \ref{sketch}. The electrode is assumed to be initially
rectangular, with its bottom surface bound to a rigid current
collector, while the electrode particles are taken to be circular
cylinders with centers located on a regular square lattice. The
mechanical evolution of the electrode is described by a model based on
Biot's theory of poroelasticity \cite{biot}, but with the additional
feature that the porous skeleton is treated as a linear viscoelastic.
Closely related models have been proposed previously for studying the
behavior and interaction of soil and groundwater in mining
applications, see {e.g.} \cite{abou,biot2}. Our primary goal here is
to use this model to predict the normal stresses, in the binder, on
the surfaces of the electrode particles and infer from these whether
(and how) delamination occurs.

The remainder of this work is as follows. In \S\ref{prior_snap}, we
formulate (and non-dimensionalize) a model for the mechanical behavior
of an electrode (estimating the sizes of the dimensionless parameters
from existing physical data). In \S\ref{the_planet}, we carry out an
ad-hoc upscaling of the governing equations to obtain homogenized
equations for the macroscopic deformations of the electrode and
exploit the small aspect ratio of the electrode (via an asymptotic
analysis) to find an approximate one-dimensional solution to this
homogenized model which we verify against a full numerical solution.
In \S\ref{time_passes}, we investigate the microscale problem around a
single electrode particle imposing boundary data compatible with the
solution to the macroscopic model. The forcing for this microscale
problem arises from: (i) increases in binder volume as it absorbs
electrolyte, and; (ii) from electrode particle volume changes in
response to cell cycling. These two forcing protocols lead to
markedly different forms of deformation corresponding to
distinct microstructural damage which can be
observed in real imaging data such as that shown in Figure
\ref{example_slices}.  Finally, in section \S\ref{discandconc} we draw
our conclusions.

\section{\label{prior_snap}Problem formulation}
Any model capable of predicting the mechanical evolution of a
composite electrode should, in principle, be multiphase --- capturing
the deformations of the polymer binder and the electrode particles, as
well as the flow of the electrolyte through the pores within the
binder. Here we assume that electrode particles can be treated as
rigid bodies which change volume in response
to lithium intercalation and de-intercalation during operation of the
cell. The assumption of particle rigidity can be justified by the much
larger mechanical moduli of the active material (typically $O(1-100)$GPa \cite{qi2,cogswell,maxisch,dileo}) in comparison to the
mechanical moduli of the binder (typically $O(1)$MPa, see table \ref{numbers}). Moreover, we restrict our interest to electrode particles that can be considered isotropic, at least on lengthscales of $O(1)\mu$m, so that when the particles expand/contract they do so without changing shape.
In a typical porous
electrode microstructure, the flow of the fluid (electrolyte) through
the pore space exerts negligibly small stresses on the porous
viscoelastic medium through which it flows (the binder). This means
that, to a good approximation, the deformation of the binder matrix
decouples from the flow of the electrolyte which allows us to
determine the mechanical state of the binder without tracking fluid
flow. Throughout we assume infinitesimal strain. 

\subsection{\label{geometry}Geometry}
A sketch of the model geometry considered is shown in Figure
\ref{sketch}. The electrode is comprised of an array of $2M+1 \times
N$ square `unit' tiles (whose sides have length $\Delta$ and
  where $M,N \in \mathbb{N}$), each of which contains a rigid circle
of active electrode material of radius $r_0^*$ at its center. We
define the half-width and thickness of the whole electrode to be $L=
(M+1/2)\Delta$ and $h=N \Delta$ respectively. Denoting the spatial
coordinates by $(x_1^*,x_2^*)$, where we adopt the convention of
denoting dimensional quantities by a star, we define the following
external boundary segments, illustrated in Figure \ref{sketch}:
\begin{eqnarray}
& \Gamma_1^* & =  \{(x_1^*,x_2^*) \, |\, -L \le x_1^* \le L,\ x_2^*=0\}, \qquad \Gamma_2^*
               =  \{(x_1^*,x_2^*) \, |\, x_1^*=L,\ 0<x_2^*<h\},\\ 
& \Gamma_3^* & =  \{(x_1^*,x_2^*) \, |\, -L \le x_1^* \le L,\  x_2^*=h\}, \qquad
               \Gamma_4^* =  \{(x_1^*,x_2^*) \, |\, x_1^*=-L,\ 0<x_2^*<h\}. 
\end{eqnarray}
The location of the $2M+1 \times N$ circular internal boundary
segments, where the active electrode particles meet the porous binder
and electrolyte, can be parametrized as follows:
\begin{equation}
\Gamma_5^{*m,n} = \{ (x_1^*,x_2^*) \, | \, (x_1^*-x_{1,m}^*)^2 +
(x_2^*-x_{2,n}^*)^2 = r_0^{*2} \} \quad \mbox{where} \quad x_{1,m}^* =
m \Delta, \, \, x_{2,n}^* = (n-1/2)\Delta, 
\end{equation}
for $m=-M,...,M$ and $n=1,...,N$. For convenience, we also introduce
the following shorthand for the complete collection of all of these
internal boundaries
\begin{equation}
\Gamma_5^* = \cup_{m=-M}^{m=M} \cup_{n=1}^{n=N} \Gamma_5^{*m,n}.
\end{equation}

\subsection{Governing equations for the porous binder}
A suitable governing equation for the deformation of the solid matrix,
in $\Omega^*$ (see Figure \ref{sketch}), is found by considering a
balance of forces on an representative elementary volume (REV) of the
porous medium (small by comparison to $\Delta$, but large by
comparison to a typical pore diameter). On neglecting inertial effects
(justified by the large time scale for electrochemical cycling, $\tau
\sim 10$hrs) and body forces we arrive at
\begin{equation} \label{start}
\frac{\partial \sigma_{ij}^*}{\partial x_i^*} = \frac{\partial p^*}{\partial x_j^*},
\end{equation}
where $p^*$ is the fluid pressure and $\sigma_{ij}^*$ are the
  components of the (symmetric) stress tensor with the indices $i$
  and $j$ running over the values $1$ and $2$, and we use the
Einstein summation convention for repeated indices. Since we work with
infinitesimal strain theory, it is not important to distinguish
between Eulerian and Lagrangian coordinates --- in fact, with small
strain theory, these two systems coincide.  The components of
  the symmetric Cauchy (infinitesimal) strain tensor
$\epsilon_{ij}^*$ (again, for a REV of the porous composite)
are defined in terms of the components of the deformation
vector, $u_i^*$, by
\begin{equation} \label{number6}
\epsilon_{ij}^* = \frac{1}{2} \left( \frac{\partial u_i^*}{\partial x_j^*} + \frac{\partial u_j^*}{\partial x_i^*} \right).
\end{equation} 
In order to define constitutive equations it is helpful to decompose
both stress and strain tensors into their volumetric and deviatoric
parts as follows: 
\begin{eqnarray}
s_{ij}^* = \sigma_{ij}^* - \delta_{ij} S^*, \quad \mbox{where} \quad S^* = \frac{1}{2} \sigma_{kk}^*,\\
e_{ij}^* = \epsilon_{ij}^* - \delta_{ij} E^*, \quad \mbox{where} \quad E^* = \frac{1}{2} \epsilon_{kk}^*.
\end{eqnarray}
Here, $s_{ij}^*$ and $e_{ij}^*$ are the components of the
deviatoric (traceless) stress and strain tensors (respectively), $S^*$
and $E^*$ are the volumetric parts of the stress and strain
(respectively). Polymer binder materials have viscoelastic properties
\cite{vino} and although there is evidence that some binders behave
nonlinearly we choose here to describe their properties using the
standard linear model of viscoelasticity (SLM) \cite{classic,zener}.
The resulting constitutive equations are
\begin{eqnarray} \label{end}
G_\tau^* \frac{\partial s_{ij}^*}{\partial t^*} + s_{ij}^* = G_2^* G_\tau^* \frac{\partial e_{ij}^*}{\partial t^*} + G_1^* e_{ij}^*,\\
K_\tau^* \frac{\partial S^*}{\partial t^*} + S^* = K_2^* K_\tau^* \frac{\partial}{\partial t^*} (E^*-\beta_{\mathrm{abs}}^*) + K_1^* (E^*-\beta_{\mathrm{abs}}^*), \label{end2}
\end{eqnarray}
where, $G_k^*$ and $K_k^*$ (for $k=\tau,1,2$) are material constants
associated with the shear and bulk deformations of the material
respectively. In particular, those with the subscripts 1 and 2
are the viscoelastic moduli (with dimensions of pressure), while those
with the subscript $\tau$ are the relaxation time scales. Notably,
these material constants are \emph{drained} coefficients --- the term
``drained'' being used to highlight the fact that these
quantities are measured by subjecting a REV of the porous medium to an
external load and measuring the resulting strains (as functions of
time) whilst allowing the fluid to freely drain out of or enter in to
the medium at constant pressure. Finally, the function
$\beta_{\mathrm{abs}}^*$ is the volumetric expansion of the binder due
to absorption of the electrolyte which is taken to be a function of
time only owing to the (roughly) uniform distributions of the polymer
and electrolyte throughout the electrode. It should be also
  noted that we do not use convective time derivatives (upper
convective, Helmholtz or other) in our formulation of the SLM
(\ref{end})-(\ref{end2}) since we are working with infinitesimal
strains.

\subsection{Boundary and initial conditions for the porous binder} 
Zero displacement (no slip) conditions are imposed on the
interface between the electrode and CC, i.e.~on $\Gamma_1^*$, see
Figure \ref{sketch}. Although it is not obvious that the electrode
should remain adhered to the CC, the images shown in Figure
\ref{example_slices} indicate that there is little or no motion there
--- indentations in the CC, resulting from the pressure applied
to the electrode during manufacture, remain close to their associated
electrode particles even after cycling. On the lateral
extremities of the electrode, $\Gamma_2^*$ and $\Gamma_4^*$, zero
stress conditions are imposed. The rationale for these conditions is
simply that the adjacent material (the electrolyte) is passive, and
therefore there is no means to provide any load/traction to these
surfaces, see Figure \ref{coin_cell}.1. The boundary conditions that
should be applied on the upper surface of the electrode,
$\Gamma_3^*$, are less obvious. Here, the electrode is in contact with
the separator, which is in turn in contact with the counter electrode,
its CC and finally (in the case of a coin cell) a spring which is used
to ensure contact is maintained between the electrochemical
components.  The stiffness of this spring is very small compared to
the moduli of the polymer binder, so that it is unable to cause any
appreciable deformation, and furthermore the separator, being composed
of a fibrous material or a flexible plastic film, is incapable of
supporting significant shear stresses. We therefore apply stress-free
conditions on the electrode's upper surface $\Gamma_3^*$. Finally, we
assume that both: (i) the state of charge of the electrode, and by
extension the volume (or the volumetric expansion) of the electrode
particles, and; (ii) the degree of electrolyte absorption, are known
\emph{a priori} and we therefore apply Dirichlet conditions on the
deformation on the electrode particle surfaces, $\Gamma_5^*$, and take
$\beta_{\mathrm{abs}}$ (the volumetric expansion of the binder due to
absorption of electrolyte) to be a known function of time. We remark
that even though non-zero deformations are permissible on some of the
boundary segments, since we are working within infinitesimal
deformation theory, the model does not constitute a free boundary
problem and the boundary conditions are applied at fixed
locations in the spatial coordinates $x_i$. In summary, the boundary
conditions on the porous skeleton are \vspace{0em}
\noindent\begin{tabularx}{\textwidth}{@{}XXX@{}}
\begin{equation} \mbox{on} \, \, \Gamma_1^*: \nonumber \end{equation} & \begin{equation} u_1^*=0  \end{equation} & \begin{equation} u_2^* = 0, \end{equation}
\end{tabularx}

\vspace{-4em}
\noindent\begin{tabularx}{\textwidth}{@{}XXX@{}}
\begin{equation} \mbox{on} \, \, \Gamma_2^*: \nonumber \end{equation} & \begin{equation} \sigma_{11}^*=0  \end{equation} & \begin{equation} \sigma_{12}^* = 0, \end{equation}
\end{tabularx}

\vspace{-4em}
\noindent\begin{tabularx}{\textwidth}{@{}XXX@{}}
\begin{equation} \mbox{on} \, \, \Gamma_3^*: \nonumber \end{equation} & \begin{equation} \sigma_{22}^*=0  \end{equation} & \begin{equation} \sigma_{12}^* = 0, \end{equation}
\end{tabularx}

\vspace{-4em}
\noindent\begin{tabularx}{\textwidth}{@{}XXX@{}}
\begin{equation} \mbox{on} \, \, \Gamma_4^*: \nonumber \end{equation} & \begin{equation} \sigma_{11}^*=0  \end{equation} & \begin{equation} \sigma_{12}^* = 0, \end{equation}
\end{tabularx}

\vspace{-4em}
\noindent\begin{tabularx}{\textwidth}{@{}XXX@{}}
\begin{equation} \mbox{on} \, \, \Gamma_5^{*m,n}:
  \nonumber \end{equation} & \begin{equation} \lefteqn{u_1^*= v_{1,m,n}^*(t) +(x_1^*-x_{1,m}^*) g^*(t^*),}\hspace{2cm}  \label{hello_giles1} \end{equation}&
\end{tabularx}

\vspace{-4em}
\noindent\begin{tabularx}{\textwidth}{@{}XXX@{}}
\begin{equation} \mbox{on} \, \, \Gamma_5^{*m,n}:
  \nonumber \end{equation} &   \begin{equation} \label{hello_giles2}\lefteqn{ u_2^* = v_{2,m,n}^*(t) +(x_2^*-x_{2,n}^*) g^*(t^*). }\hspace{2cm} \end{equation}&
\end{tabularx}

Here the final two conditions represent the uniform growth of the circular
electrode particles from an initial radius $r_0^*$ to a radius
$r_0^*(1+g^*(t^*))$, and allow for a displacement $v^*_{i,m,n}(t^*)$ of the centre of
the $(m,n)$-th cylinder. This displacement is determined by imposing that
there is no net force on the cylinder, so that\footnote{If we did not
  allow for the displacement $v_{i,m,n}(t^*)$ and fixed the position
  of the cylinders, then they would apply a non-zero net
  force to the binder. This is equivalent to imagining the ends of the
  cylinders to be fastened to some support, rather than allowing them
  to move freely.} 
\begin{equation}
 \int_{\Gamma_5^{*m,n}} \sigma^*_{ij} n_j \, \d s= 0,
\end{equation}
where $(n_1,n_2)$ is the normal to $\Gamma_5^{*m,n}$.
In principle, we should also allow a (linearised) rotation of the
cylinder, determined by the condition that there is no net
torque. However, we will see later that the torque induced on the
binder by ignoring such rotations is small. 

To close the solid-state component of the model, it remains to specify
initial data for the relevant quantities. We assume initially, prior
to the addition of the electrolyte, that the electrode is in a zero
stress state so that\\
\vspace{0em}
\noindent\begin{tabularx}{\textwidth}{@{}XXX@{}}
\begin{equation} \mbox{at} \, \, t^*=0 \nonumber \end{equation} & \begin{equation} u_i^*|_{t^*=0} = 0 \end{equation} & \begin{equation} \sigma_{ij}^*|_{t^*=0} = 0. \label{ICSS} \end{equation} 
\end{tabularx}

\subsection{\label{fluid_bit}Governing equations for the electrolyte}
It is usual to relate the flux of a fluid through a porous medium to
the pressure gradient within the fluid via Darcy's Law which, for a
deformable porous medium takes the form \cite{fowler}
\begin{equation}
\phi_f^* \left( w_i^* - \frac{\partial u_i^*}{\partial t^*} \right) = - \frac{k}{\mu} \frac{\partial p^*}{\partial x_i^*}, \label{Peq}
\end{equation}
where $\phi_f^*$, $w_i^*$, $k$, $\mu$, are the volume fraction of
fluid, a component of the fluid velocity, the permeability of the
porous medium and the fluid viscosity. The problem for the fluid
component is closed by making further statements on the conservation
of mass of each phase (the fluid and the solid skeleton). On noting
that, by definition, $\phi_f^*+\phi_s^*=1$, where $\phi_s^*$ and
$\phi_f^*$ are the volume fractions of the solid and fluid phases
respectively, these conservation equations are
\begin{align}
\frac{\partial}{\partial t^*} \left(  \phi_f^* \right) + \frac{\partial}{\partial x_i^*} \left( w_i^*  \phi_f^* \right) =0,\\
\frac{\partial}{\partial t^*} \left( 1- \phi_f^* \right) + \frac{\partial}{\partial x_i^*} \left( \frac{\partial u_i^*}{\partial t^*} (1-\phi_f^*) \right) =0.
\end{align} 
Summing the equations above and substituting for $\phi_f^* \left(
  w_i^* - {\partial u_i^*}/{\partial t^*} \right)$ from (\ref{Peq}),
in the standard fashion, gives 
\begin{equation*}
\frac{\dd^2 u_i^*}{\dd t^* \dd
  x_i^*} = \frac{k}{\mu} \frac{\dd^2 p^*}{\dd x_i^{*2}}.  
\end{equation*}
The typical pressure scale $P_0$ can be estimated from this equation in
terms of typical time, displacement and length scales $\tau$, $U_0$
and $L$, respectively. It is given by 
\begin{equation*}
P_0=\frac{\mu L U_0}{k \tau}  
\end{equation*}
and since the flow occurs along the width of the electrode (it cannot
flow through the impermeable current collectors) we take $L$ to be the
electrode half-width, \ie $L \sim 1$cm. On referring to table
\ref{numbers}, we can compute a generous estimate of the pressure 
$P_0 \sim 10^1$Pa. In order to work out whether the fluid pressure
plays a significant part in the viscoelastic deformation of the binder,
we use our estimate for $P_0$ in the force balance equation
(\ref{start}) in order to compare the sizes of the mechanical stress
and fluid pressure gradient terms on the left- and right-hand side of
this equation. The typical size $\Sigma_0$ for the stress tensor is
obtained from the constitutive equations (\ref{end})-(\ref{end2})
using an estimate for the size of the strain tensor; on using a
lengthscale $h\sim100\mu$m (\ie the electrode thickness), this gives
\begin{equation*} 
\Sigma_0=\frac{G_1^* U_0}{h}.  
\end{equation*}
On referring to table
\ref{numbers} we obtain an estimate of the size of the stress tensor
$\Sigma_0 \sim 10^5$Pa which is a factor of $10^{4}$ larger than $P_0$
and we conclude that it is reasonable to neglect the fluid pressure
term in (\ref{start}), by replacing it by the following purely
mechanical force balance equation:
\begin{equation} \label{start2}
\frac{\partial \sigma_{ij}^*}{\partial x_i^*} = 0.
\end{equation}

\begin{table}
    \begin{center}
    \begin{tabular}{| l | l | r |}
    \hline
    Symbol 	& Description & Value \\ \hline
    $\Delta$ 	& Length of the side of a square `unit' tile & 1--10$\mu$m \\ 
    $r_0^*$ 	& Electrode particle radius & 1--10$\mu$m \cite{hanshuo,nester}\\ 
    $h$ 	& Electrode thickness & $100\mu$m \cite{hanshuo,nester}\\ 
    $L$		& Electrode half-width & $1$cm \cite{hanshuo,nester}\\
    $\beta_0$	& Volumetric expansion of the binder due to electrolyte absorption & 0.4--0.6 \cite{chen,yash,park} \\
    $\tau$	& Timescale for cell (dis-)charge & $O(10)$hrs \cite{rdp} \\
    $\mu$	& Electrolyte viscosity (EC/DMC with 1 molar LiPF$_6$) & 2.6$\times$10$^{-3}$Pa$\cdot$s \cite{sigald} \\
    $k$		& Electrode permeability (or hydraulic conductivity) & $O(10^{-15})$m$^2$ \\
    $G_1^*$	& Drained viscoelastic shear modulus & $O(10^6)$Pa \cite{xiaosong2}\\
    $G_2^*$	& Drained viscoelastic shear modulus & $O(10^6)$Pa \cite{xiaosong2} \\
    $G_\tau^*$	& Drained shear relaxation timescale & $O(0.1)$hrs \cite{xiaosong2} \\
    $K_1^*$	& Drained viscoelastic bulk modulus & $O(10^6)$Pa \cite{xiaosong2} \\
    $K_2^*$	& Drained viscoelastic bulk modulus & $O(10^6)$Pa \cite{xiaosong2} \\
    $K_\tau^*$	& Drained bulk relaxation timescale & $O(0.1)$hrs \cite{xiaosong2} \\
    \hline
    $G_\tau$	& Ratio of timescales for shear relaxation and cell (dis-)charge &$O(10^{-2})$\\	
    $G_2$	& Ratio of the second and first viscoelastic shear moduli &$O(1)$\\
    $K_\tau$	& Ratio of timescale for bulk relaxation and cell (dis-)charge &$O(10^{-2})$\\
    $K_1$	& Ratio of the first bulk modulus to the first shear modulus &$O(1)$\\
    $K_2$	& Ratio of the second bulk modulus to the first shear modulus&$O(1)$\\
    $\gamma$	& Ratio of the electrode thickness and width &$O(10^{-2})$\\
    $\lambda$	& Ratio of the length of a `unit tile' to electrode thickness &$O(10^{-1}-10^{-2})$\\
    $r_0$	& Ratio of the radius of an electrode particle to length of a `unit tile' &$O(10^{-1}-10^{-2})$\\
    \hline
    \end{tabular}
    \caption{\label{numbers}Descriptions and estimates of both the dimensional (upper portion) and dimensionless parameters (lower portion).}
    \end{center}
\end{table}

\subsection{\label{scalings}Non-dimensionalization}
We non-dimensionalize the model by setting:
\begin{align} \label{puppet1}
x_i^* &= h x_i, & t^* &= \tau t, & x_{i,m}^* &= h x_{i,m}\\
\beta_{\mathrm{abs}}^* &= \beta_0 \beta_{\mathrm{abs}}, & u_i^* &= \beta_0 h u_i, & g^* &= \beta_0 g,\\
\epsilon_{ij}^* &= \beta_0 \epsilon_{ij}, & e_{ij}^* &= \beta_0 e_{ij}, & E^* &= \beta_0 E,\\
\label{puppets_bum} \sigma_{ij}^* &= \beta_0 G_1^* \sigma_{ij}, &
                                                                  s_{ij}^* &= \beta_0 G_1^* s_{ij}, & S^*&= \beta_0 G_1^* S\\
v^*_{i,m,n}& = \beta_0 h v_{i,m,n}, 
\end{align}
where $\beta_0$ is the typical size of volumetric expansion of the binder owing to electrolyte absorption. The non-dimensionalization leads to a system characterized by the dimensionless parameters:
\begin{eqnarray} \label{params}
G_{\tau} = \frac{G_\tau^*}{\tau}, \, \, \, 
G_2 = \frac{G_2^*}{G_1^*}, \, \, \, 
K_{\tau}^* = \frac{K_\tau^*}{\tau}, \, \, \, 
K_1 = \frac{K_1^*}{G_1^*}, \, \, \, 
K_2 = \frac{K_2^*}{G_1^*}, \, \, \, 
\gamma=\frac{h}{L}, \, \, \, 
\lambda = \frac{\Delta}{h}, \, \, \, r_0 = \frac{r_0^*}{h}.
\end{eqnarray}
Estimates of these parameters depend upon the different lengthscales
in problem ($\Delta$, $r_0^*$, $h$ and $L$), as well as the timescale
$\tau$ for cell (dis-)charge. We base these, in turn, on the devices
studied in \cite{hanshuo,nester} from which we obtain the estimates
shown in table \ref{numbers}. We base our estimate of the electrolyte
viscosity, $\mu$, on a 1 molar LiPF$_6$ solution in EC/DMC --- one of
the most common electrolytes used in both commercial and research
cells --- its value shown in table \ref{numbers} is taken from a
Sigma-Aldrich data sheet (the viscosities of most other common battery
electrolytes are similar). The electrode permeability can be estimated
from, for example, the Carman-Kozeny formula \cite{car37,koz27}. Using
suitable values for the porosity and approximations of the electrode
geometry a value of $k=O(10^{15})$m$^2$ is derived --- which is
comparable to that of sandstone. We base our estimates of $\beta_0$ on
the data provided in \cite{chen,yash,park}, where it is stated that the
densities of EC:DMC, PVDF and CMC are $1.2$g/cc, $1.8$g/cc and
$1.6$g/cc respectively. Furthermore, according to \cite{chen,yash,park}, a
typical volumetric expansion for a binder, either PVDF or CMC, is
around 75\% by weight. Thus, an estimate for $\beta_0$ is in the range
0.4--0.6. Typical values for mechanical coefficients are more
difficult to estimate. It appears that they depend strongly on not
only the polymer under consideration, but also the carbon black
content and processing conditions. The situation is further
complicated by the fact that these polymers tend to soften as they
absorb electrolyte --- contrast, {e.g.} the values given in
\cite{xiaosong2} against those in \cite{vino}. However, the relaxation
modulus arising from experimental measurements of softened (via
electrolyte absorption) PVDF has been given in \cite{xiaosong2} and a
six term Prony series was proposed. Here we have opted to model the
binder behavior using a SLM which corresponds to a two term Prony
series \cite{classic}. We therefore approximate the time dependent
modulus in \cite{xiaosong2} using $E(t) = E_{\infty} + E_0
\exp(-t/t_0)$ with $E_\infty = 0.52$MPa, $E_0 = 1.03$MPa and $t_0 =
700$s which gives rise to the estimates of $G_i^*$ and $K_i^*$ (for
$i=1,2,\tau$) given in table \ref{numbers}.


\subsection{The dimensionless problem}
Applying the scalings (\ref{puppet1})-(\ref{puppets_bum}) to equations
(\ref{start})-(\ref{hello_giles2}) leads to the following dimensionless
system for dimensionless (unstarred) variables 

\noindent\begin{tabularx}{\textwidth}{@{}XX@{}}
  \begin{equation}
 \frac{\partial \sigma_{ij}}{\partial x_j}  = 0,
\label{nd_start}
  \end{equation} &
  \begin{equation}
\epsilon_{ij} = \frac{1}{2} \left( \frac{\partial u_i}{\partial x_j} +
  \frac{\partial u_j}{\partial x_1} \right), \label{strain}
  \end{equation} 
\end{tabularx} 

\vspace{-3em}
\noindent\begin{tabularx}{\textwidth}{@{}XX@{}}
  \begin{equation}
e_{ij} = \epsilon_{ij} - \delta_{ij} E,
  \end{equation}  &
  \begin{equation}
\label{ee3} s_{ij} = \sigma_{ij} - \delta_{ij} S,
  \end{equation} 
\end{tabularx}

\vspace{-3em}
\noindent\begin{tabularx}{\textwidth}{@{}XX@{}}
   \begin{equation}
E= \frac{1}{2} \epsilon_{kk},
  \end{equation}&  \begin{equation}
S=\frac{1}{2} \sigma_{kk},
  \end{equation}
\end{tabularx}

\vspace{-2em}
  \begin{equation}
\label{ee4} G_\tau \frac{\partial s_{ij}}{\partial t} + s_{ij} = G_2 G_\tau \frac{\partial e_{ij}}{\partial t} + e_{ij},
  \end{equation} 
\vspace{-2em}
\begin{equation}
K_\tau \frac{\partial S}{\partial t} + S = K_2 K_\tau \frac{\partial}{\partial t} (E-\beta_{\mathrm{abs}}) + K_1 (E-\beta_{\mathrm{abs}}) \label{nd_eq_end}
  \end{equation} 
with boundary and initial conditions 

\vspace{1em}
\noindent\begin{tabularx}{\textwidth}{@{}cXX@{}}
on $\Gamma_1$:   
& \vspace{-3em}\begin{equation} u_1=0  \end{equation} & \vspace{-3em}\begin{equation} u_2 = 0, \end{equation}
\end{tabularx}

\vspace{-1.5em}
\noindent\begin{tabularx}{\textwidth}{@{}cXX@{}}
on $\Gamma_2$:   
& \vspace{-3em}\begin{equation} \sigma_{11}=0  \end{equation} &
\vspace{-3em}
\begin{equation} \sigma_{12} = 0, \end{equation}
\end{tabularx}

\vspace{-1.5em}
\noindent\begin{tabularx}{\textwidth}{@{}cXX@{}}
on $\Gamma_3$: 
& \vspace{-3em}\begin{equation} \sigma_{22}=0  \end{equation} &
\vspace{-3em}\begin{equation} 
  \sigma_{12} = 0, \end{equation} 
\end{tabularx}

\vspace{-1.5em}
\noindent\begin{tabularx}{\textwidth}{@{}cXX@{}} 
on $\Gamma_4$: 
& \vspace{-3em}\begin{equation} \sigma_{11}=0  \end{equation} &\vspace{-3em} \begin{equation}
  \sigma_{12} = 0, \end{equation} 
\end{tabularx}

\vspace{-1.5em}
\noindent\begin{tabularx}{\textwidth}{@{}cXX@{}} 
on $\Gamma_5$: 
&\vspace{-3em} \begin{equation} u_1= v_{1,m,n}(t) + (x_1-x_{1,m}) g(t)
  \end{equation} 
& \vspace{-3em}\begin{equation}   u_2 =  v_{2,m,n}(t) +(x_2-x_{2,n}) g(t),\end{equation}
\end{tabularx}

\vspace{-1.5em}
\noindent\begin{tabularx}{\textwidth}{@{}cXX@{}}
               at $t=0$:  &\vspace{-3em}\begin{equation} u_i|_{t=0} =0  \end{equation} & \vspace{-3em}\begin{equation}  \sigma_{ij}|_{t=0} = 0  \end{equation}
\end{tabularx}
and the additional constraint
\begin{equation}
 \int_{\Gamma_5^{m,n}} \sigma_{ij} n_j \, \d s= 0.\label{smell_my_cheese}
\end{equation}

\noindent
In dimensionless variables, the locations of the various boundary segments are 
\begin{eqnarray}
& \label{diddly}\qquad  \Gamma_1 & =  \{(x_1,x_2) \, |\,
                            -1/\gamma<x_1<1/\gamma,\ x_2=0\}, \qquad
                            \Gamma_2 =  \{(x_1,x_2) \, |\,
                            x_1=1/\gamma,\ 0<x_2<1\},\\ 
& \label{dodo} \qquad\Gamma_3 & =  \{(x_1,x_2) \, |\,
                          -1/\gamma<x_1<1/\gamma,\ 
                          x_2=1\}, \qquad \Gamma_4 =  \{(x_1,x_2) \, |\,
                          x_1=-1/\gamma,\ 0<x_2<1\},\\ 
& \quad \qquad\Gamma_5^{m,n} & = \{ (x_1,x_2) \, | \, (x_1-x_{1,m})^2 +
                   (x_2-x_{2,n})^2 = r_0^2 \}, \, \, \, \, \Gamma_5 =
                   \cup_{m=-M}^{m=M} \cup_{n=1}^{n=N}
                   \Gamma_5^{m,n},
\end{eqnarray}
where $x_{1,m} = m \lambda$, $x_{2,n} =
                (n-1/2)\lambda$ for $m=-M,\dots,M$ and 
                $n=1,\dots,N$.

\section{\label{the_planet}Upscaling and the macroscale problem}

There are two predominant scales in this problem: there is a
microscale $O(1-10\mu \mbox{m})$ defined by the typical distance
between electrode particles and a macroscale defined by the dimensions
of the electrode\footnote{In fact it could be argued that the
  macroscale is subdivided into a mesoscale of $O(100\mu \mbox{m})$,
  defined by the electrode height, and a genuine macroscale of $O(1
  \mbox{cm})$, defined by the electrode width. However, for the
  current purposes it proves convenient to incorporate both these
  scales into a single macroscale.} $O(1\mbox{mm})$. On the microscale
we consider the local deformation of the binder matrix around
individual particles as a result of either: (i) the binder
swelling in response to soaking-up electrolyte, or, (ii) the
electrode particles expanding and contracting in response to
lithiation and delithiation; whereas on the macroscale we consider the
bulk response of the entire material, including binder and electrode
particles. A proper treatment of the macroscale problem
requires homogenization of the microscale equations to obtain
effective medium equations approximating the macroscopic behavior of
the electrode (see for example \cite{rc,rdp}). Here however we are
primarily interested in the results of the microscale problem because
we wish to find the microscopic stress distribution around an
electrode particle in order to see whether, and if so how,
delamination of the binder from the electrode particle occurs. In
order to accomplish this we consider a locally periodic array of
electrode particles and solve the problem in a single periodic
`unit' cell around one of these particles. Nevertheless, the
imposition of appropriate boundary conditions on the edge of this cell
still requires knowledge of the macroscopic solution. In particular,
we seek to demonstrate that, because of its large aspect ratio, the
macroscopic deformation of the composite material is one-dimensional,
occurring in the direction normal to the current collector
$\Gamma_1$, except close to the edges of the electrode. Since the
information we require from the macroscale is limited, we look to
accomplish this without resorting to a lengthy homogenization
procedure. Instead we use physical intuition to write down a
plausible set of macroscale equations, and leave the task of
systematically determining the macroscopic parameters in terms
of the microscale geometry and parameters to a further work.

\begin{figure}          \centering
\includegraphics[width=0.5\textwidth]{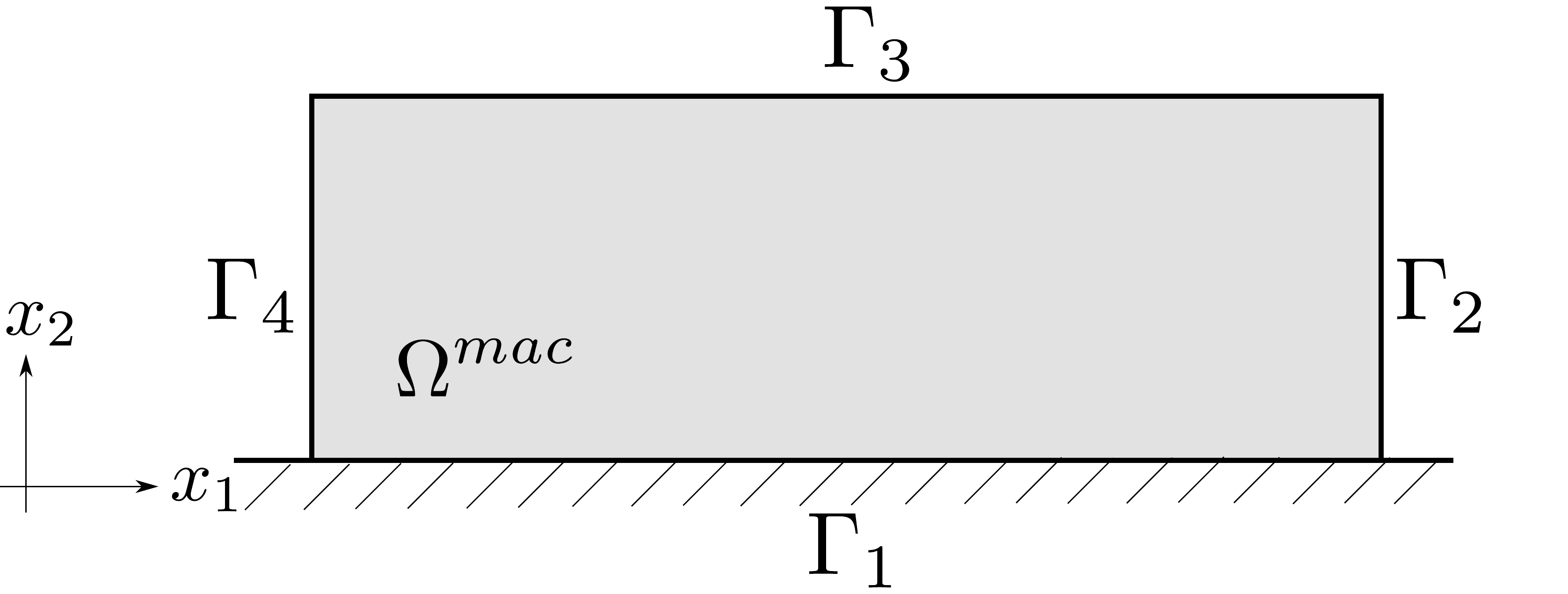}
\caption{Schematic of the geometry of the macroscale problem.}
\label{macro_scheme}
\end{figure}

The simplified homogeneous geometry for the macroscale problem
  is shown in Figure \ref{macro_scheme}. We note that the locations of
  the external boundaries of the electrode are unchanged by the
  upscaling procedure. Thus, we retain the notation for the external
  boundary segments from the full problem. Explicit definitions of
  $\Gamma_k$ for $k=1,\dots,4$ are given in
  (\ref{diddly})-(\ref{dodo}).

It is clear that the effective stress and strain at the macroscale,
denoted by the superscript ``eff'', will still satisfy
(\ref{nd_start})-(\ref{strain}), and we can define effective
deviatoric and volumetric stresses as usual, so that

\noindent\begin{tabularx}{\textwidth}{@{}XX@{}}
  \begin{equation}
 \frac{\partial \sigma_{ij}^{\mathrm{eff}}}{\partial x_j}  = 0,
\label{macro_start}
  \end{equation} &
 \begin{equation}
\epsilon_{ij}^{\mathrm{eff}} = \frac{1}{2} \left( \frac{\partial
    u_i^{\mathrm{eff}}}{\partial x_j} + \frac{\partial u_j^{\mathrm{eff}}}{\partial x_i}
\right),
  \end{equation}  
\end{tabularx}

\vspace{-3em}
\noindent\begin{tabularx}{\textwidth}{@{}XX@{}}
  \begin{equation}
s_{ij}^{\mathrm{eff}} = \sigma_{ij}^{\mathrm{eff}} - \delta_{ij} S^{\mathrm{eff}},
  \end{equation}  & \begin{equation}
S^{\mathrm{eff}}=\frac{1}{2} s_{kk}^{\mathrm{eff}}, \label{snaf2}
  \end{equation} 
\end{tabularx}

\vspace{-3em}
\noindent\begin{tabularx}{\textwidth}{@{}XX@{}}
  \begin{equation} \label{sharon}
e_{ij}^{\mathrm{eff}} = \epsilon_{ij}^{\mathrm{eff}} - \delta_{ij} E^{\mathrm{eff}},
  \end{equation} &  \begin{equation} \label{jack}
E^{\mathrm{eff}}= \frac{1}{2} \epsilon_{kk}^{\mathrm{eff}}.
  \end{equation} 
\end{tabularx}

What is less clear is the constitutive equation which should hold for
the homogenised material. There is no particular reason it should
correspond to a SLM (or a two term Prony
series) --- more likely there will be a continuum of timescales
involved. Nevertheless, for the small amount of information we
require, we expect that approximating the macroscopic constitutive
behaviour by a SLM with effective coefficients is sufficient.

The final part of the homogenised model concerns the macroscopic
effect of the changing volumes of the electrode particles on the
microscopic length scale. This is captured by an ``effective'' growth
function $\beta_{\mathrm{am}}^{\mathrm{eff}}$ --- given by the ratio of the volume
expansion of an electrode particle to the volume the periodic cell in
which it is embedded --- which appears in the equations in the same way
that the binder swelling does.  Notably, because we assume that
particles are uniformly distributed in space, $\beta_{\mathrm{am}}^{\mathrm{eff}}(t)$
is a function of time only.  Thus the remaining macroscale equations
(to be solved on
$\Omega^{\mathrm{mac}}$, see Figure \ref{macro_scheme}) are given by: \\
\noindent\begin{tabularx}{\textwidth}{@{}XX@{}}
  \begin{equation}
G^{\mathrm{eff}}_\tau \frac{\partial s_{ij}^{\mathrm{eff}}}{\partial
  t} + s_{ij}^{\mathrm{eff}} = G^{\mathrm{eff}}_2 G^{\mathrm{eff}}_\tau \frac{\partial 
  e_{ij}^{\mathrm{eff}}}{\partial t} + e_{ij}^{\mathrm{eff}}, 
\label{eq:Geff}
  \end{equation} 
\end{tabularx}

\vspace{-2em}
  \begin{equation}
K^{\mathrm{eff}}_\tau \frac{\partial S^{\mathrm{eff}}}{\partial t} + S^{\mathrm{eff}} =
K^{\mathrm{eff}}_2 K^{\mathrm{eff}}_\tau \frac{\partial}{\partial t} \left[ E^{\mathrm{eff}}-\left(
    \beta_{\mathrm{abs}}^{\mathrm{eff}} + \beta_{\mathrm{am}}^{\mathrm{eff}}  \right) \right] +
K^{\mathrm{eff}}_1 \left[ E^{\mathrm{eff}} - \left( \beta_{\mathrm{abs}}^{\mathrm{eff}} +
    \beta_{\mathrm{am}}^{\mathrm{eff}} \right) \right].
\label{eq:Keff}
  \end{equation} 

The boundary conditions on the outer edges of the electrode remain
unchanged, and the conditions on the interface between the electrode
and the embedded electrode particles are no longer needed. The boundary
conditions that close the macroscopic problem are therefore 

\noindent\begin{tabularx}{\textwidth}{@{}XXX@{}}
\begin{equation} \mbox{on} \, \, \Gamma_1: \nonumber \end{equation} & \begin{equation} u_1^{\mathrm{eff}}=0  \label{bob} \end{equation} & \begin{equation} u_2^{\mathrm{eff}} = 0, \label{kelly} \end{equation}
\end{tabularx}

\vspace{-4em}
\noindent\begin{tabularx}{\textwidth}{@{}XXX@{}}
\begin{equation} \mbox{on} \, \, \Gamma_2: \nonumber \end{equation} & \begin{equation} \sigma_{11}^{\mathrm{eff}}=0  \end{equation} & \begin{equation} \sigma_{12}^{\mathrm{eff}} = 0, \end{equation}
\end{tabularx}

\vspace{-4em}
\noindent\begin{tabularx}{\textwidth}{@{}XXX@{}}
\begin{equation} \mbox{on} \, \, \Gamma_3: \nonumber \end{equation} & \begin{equation} \sigma_{22}^{\mathrm{eff}}=0 \label{barf1} \end{equation} & \begin{equation} \label{barf2} \sigma_{12}^{\mathrm{eff}} = 0, \end{equation}
\end{tabularx}

\vspace{-4em}
\noindent\begin{tabularx}{\textwidth}{@{}XXX@{}}
\begin{equation} \mbox{on} \, \, \Gamma_4: \nonumber \end{equation}
& \begin{equation} \sigma_{11}^{\mathrm{eff}}=0  \end{equation}
& \begin{equation}  \sigma_{12}^{\mathrm{eff}} = 0, \end{equation} 
\end{tabularx}

\vspace{-4em}
\noindent\begin{tabularx}{\textwidth}{@{}XXX@{}}
\begin{equation} \mbox{at} \, \, t=0: \nonumber \end{equation} & \begin{equation} u_{i}^{\mathrm{eff}}|_{t=0}=0  \end{equation} & \begin{equation} \label{end_macro_bcs} \sigma_{ij}^{\mathrm{eff}}|_{t=0} = 0. \end{equation}
\end{tabularx}

We seek an asymptotic solution to
(\ref{macro_start})-(\ref{end_macro_bcs}), valid throughout the bulk
of the electrode (away from the lateral boundaries $\Gamma_2$ and
$\Gamma_4$) for aspect ratio $\gamma \ll 1$.  We rescale $x_1$ with
$1/\gamma$ and take the limit $\gamma \ra 0$.  On taking the
leading-order terms, integrating with respect to $x_2$ and imposing
the boundary conditions (\ref{barf1}) and (\ref{barf2}) we find that
$\sigma^{\mathrm{eff}}_{12}=\sigma^{\mathrm{eff}}_{22}=0$. Thus
\begin{eqnarray} \label{ozzy}
\sigma^{\mathrm{eff}} = \left( \begin{array}{cc}
  \sigma_{11}^{\mathrm{eff}} & 0 \\*[3mm] 0 & 0 \end{array} \right),
\qquad S^{\mathrm{eff}} = \frac{\sigma_{11}^{\mathrm{eff}}}{2},
\qquad s = \left( \begin{array}{cc} \sigma_{11}^{\mathrm{eff}}/2 &
  0 \\ 0 & -\sigma_{11}^{\mathrm{eff}}/2 \end{array} \right). 
\end{eqnarray} 
Similarly, at leading order,
\begin{eqnarray} \label{gus}
\quad\epsilon^{\mathrm{eff}} = \left( \begin{array}{cc} 0 & \displaystyle
  \frac{1}{2} \frac{\partial u_{1}^{\mathrm{eff}}}{\partial x_2}
  \\*[3mm] \displaystyle \frac{1}{2} \frac{\partial
    u_{1}^{\mathrm{eff}}}{\partial x_2} & \displaystyle
  \frac{\partial u_{2}^{\mathrm{eff}}}{\partial x_2} \end{array}
\right), \qquad  E^{\mathrm{eff}} = \frac{1}{2} \frac{\partial
  u_{2}^{\mathrm{eff}}}{\partial x_2}, \qquad e^{\mathrm{eff}} =
\left( \begin{array}{cc}  \displaystyle - \frac{1}{2}\frac{\partial
    u_{2}^{\mathrm{eff}}}{\partial x_2} & \displaystyle \frac{1}{2}
  \frac{\partial u_{1}^{\mathrm{eff}}}{\partial x_2} \\*[3mm]
  \displaystyle \frac{1}{2} \frac{\partial
    u_{1}^{\mathrm{eff}}}{\partial x_2} & \displaystyle \frac{1}{2}
  \frac{\partial u_{2}^{\mathrm{eff}}}{\partial x_2} \end{array}
\right). 
\end{eqnarray}
Substituting (\ref{ozzy}c) and (\ref{gus}c) into the $12$ component of
\eqref{eq:Geff} gives
\begin{equation}
G_2^{\mathrm{eff}} G_\tau^{\mathrm{eff}} \frac{\partial}{\partial t}
\left( \frac{\partial 
  u_{1}^{\mathrm{eff}}}{\partial x_2} \right) + \frac{\partial
  u_{1}^{\mathrm{eff}}}{\partial x_2} = 0. 
\end{equation}
The only solution to this problem satisfying both initial data and the
boundary condition (\ref{bob}) is
\begin{equation}
u_{1}^{\mathrm{eff}} = 0. \label{deaf}
\end{equation}

A pair of coupled equations for the remaining unknowns,
$\sigma_{11}^{\mathrm{eff}}$ and $\partial
u_{2}^{\mathrm{eff}}/\partial x_2$, are derived by substituting
(\ref{ozzy}) and (\ref{gus}) into the $11$ (or $22$) component of
\eqref{eq:Geff} and \eqref{eq:Keff}; they are
\begin{eqnarray}
G^{\mathrm{eff}}_\tau \frac{\partial \sigma_{11}^{\mathrm{eff}}}{\partial t} +
\sigma_{11}^{\mathrm{eff}} + G^{\mathrm{eff}}_2 G^{\mathrm{eff}}_\tau \frac{\partial}{\partial t}
\left( \frac{\partial u_{2}^{\mathrm{eff}}}{\partial x_2} \right) +
\frac{\partial u_{2}^{\mathrm{eff}}}{\partial x_2} =0,\\ 
\quad \quad \quad K^{\mathrm{eff}}_\tau \frac{\partial
  \sigma_{11}^{\mathrm{eff}}}{\partial t} +
\sigma_{11}^{\mathrm{eff}} = K^{\mathrm{eff}}_2 K^{\mathrm{eff}}_\tau \frac{\partial}{\partial t}
\left( \frac{\partial u_{2}^{\mathrm{eff}}}{\partial x_2} -
(\beta_{\mathrm{abs}}^{\mathrm{eff}} +
\beta_{\mathrm{am}}^{\mathrm{eff}} ) \right) + K^{\mathrm{eff}}_1 \left(
\frac{\partial u_{2}^{\mathrm{eff}}}{\partial x_2} -
(\beta_{\mathrm{abs}}^{\mathrm{eff}} +
\beta_{\mathrm{am}}^{\mathrm{eff}} ) \right).  
\end{eqnarray}
On using Laplace transforms to solve the above equations, integrating
the resulting expression for $\partial u_{2}/\partial x_2$ with
respect to $x_2$, and imposing the boundary condition (\ref{kelly}) we
find  
\begin{eqnarray}
\qquad  u_{2}^{\mathrm{eff}} &=& x_2
\mathcal{L}^{-1} \left\{ 
\frac{\left(
\bar{\beta}_{\mathrm{abs}}^{\mathrm{eff}}(s) +
\bar{\beta}_{\mathrm{am}}^{\mathrm{eff}}(s) \right)(G^{\mathrm{eff}}_\tau s +1)(K^{\mathrm{eff}}_\tau
  K^{\mathrm{eff}}_2 s +K^{\mathrm{eff}}_1)}{(K^{\mathrm{eff}}_\tau
  s+1)(G^{\mathrm{eff}}_\tau G^{\mathrm{eff}}_2 s +1)+(K^{\mathrm{eff}}_\tau K^{\mathrm{eff}}_2 s
  +K^{\mathrm{eff}}_1)(G^{\mathrm{eff}}_\tau s +1)}
\right\},\label{snapper}
 \\ 
 \sigma_{11}^{\mathrm{eff}} &=& - \mathcal{L}^{-1} \left\{
\frac{\left( \bar{\beta}_{\mathrm{abs}}^{\mathrm{eff}}(s) +
\bar{\beta}_{\mathrm{am}}^{\mathrm{eff}}(s) \right)(G^{\mathrm{eff}}_\tau G^{\mathrm{eff}}_2 s +1)(K^{\mathrm{eff}}_\tau K^{\mathrm{eff}}_2 s
  +1)}{(K^{\mathrm{eff}}_\tau s+1)(G^{\mathrm{eff}}_\tau G^{\mathrm{eff}}_2 s
  +1)+(K^{\mathrm{eff}}_\tau K^{\mathrm{eff}}_2 s
  +K^{\mathrm{eff}}_1)(G^{\mathrm{eff}}_\tau s +1)}
\right\}, \label{macro_sol_end} 
\end{eqnarray}
where $\mathcal{L}^{-1}$ denotes the inverse Laplace transform.

Most importantly, we have found that the thin geometry of the
electrode and the zero displacement condition on the current collector
forces the deformation to be essentially one-dimensional (in the
$x_2$-direction) --- see (\ref{deaf}). Crucially, as we will see in the
subsequent section, the relatively simple structure of the solution on
the macroscopic scale will allow us to construct appropriate boundary
conditions to apply at the microscopic scale.

\begin{figure}          \centering
\includegraphics[width=0.4\textwidth]{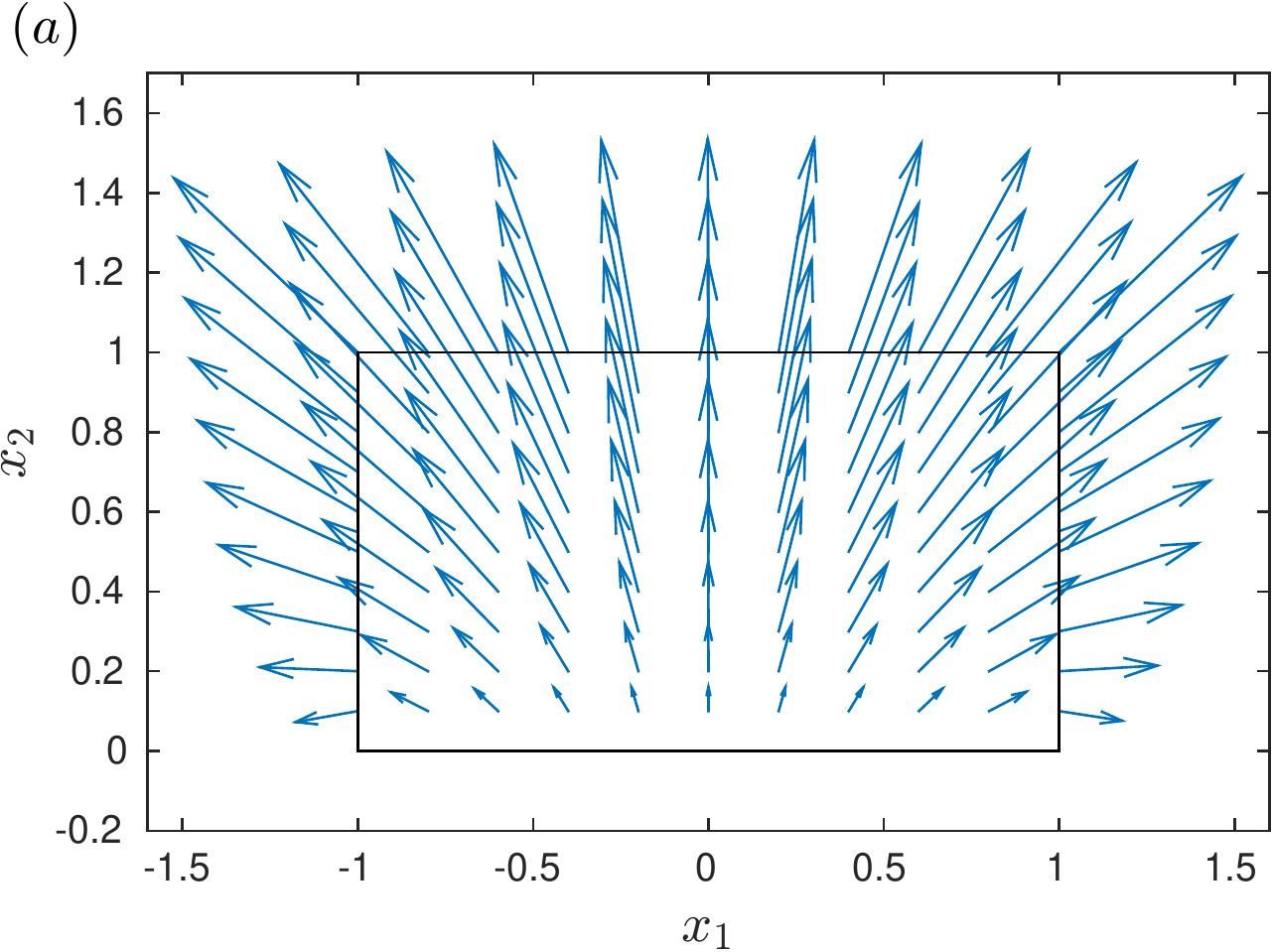}
\includegraphics[width=0.4\textwidth]{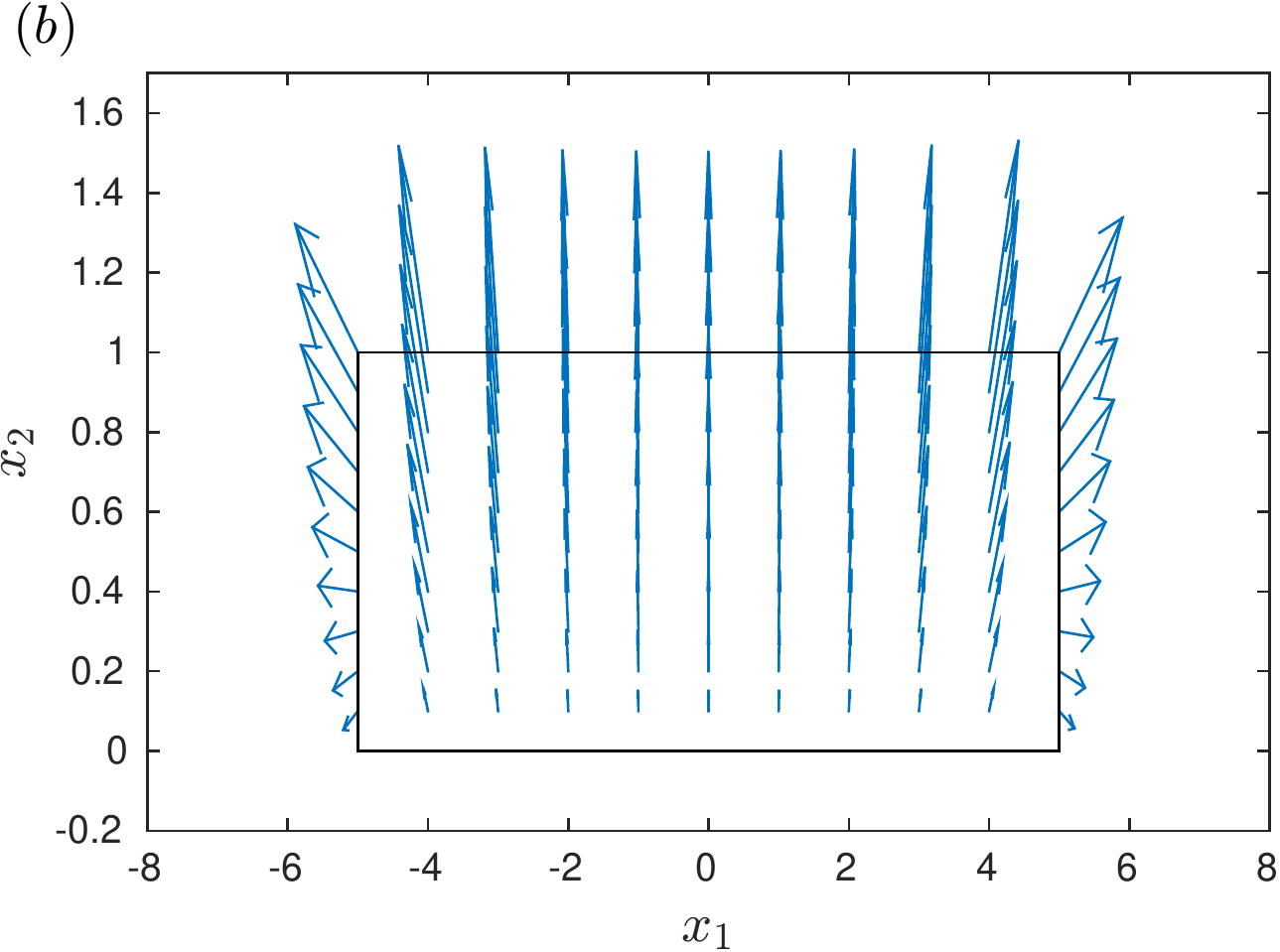}
\includegraphics[width=0.4\textwidth]{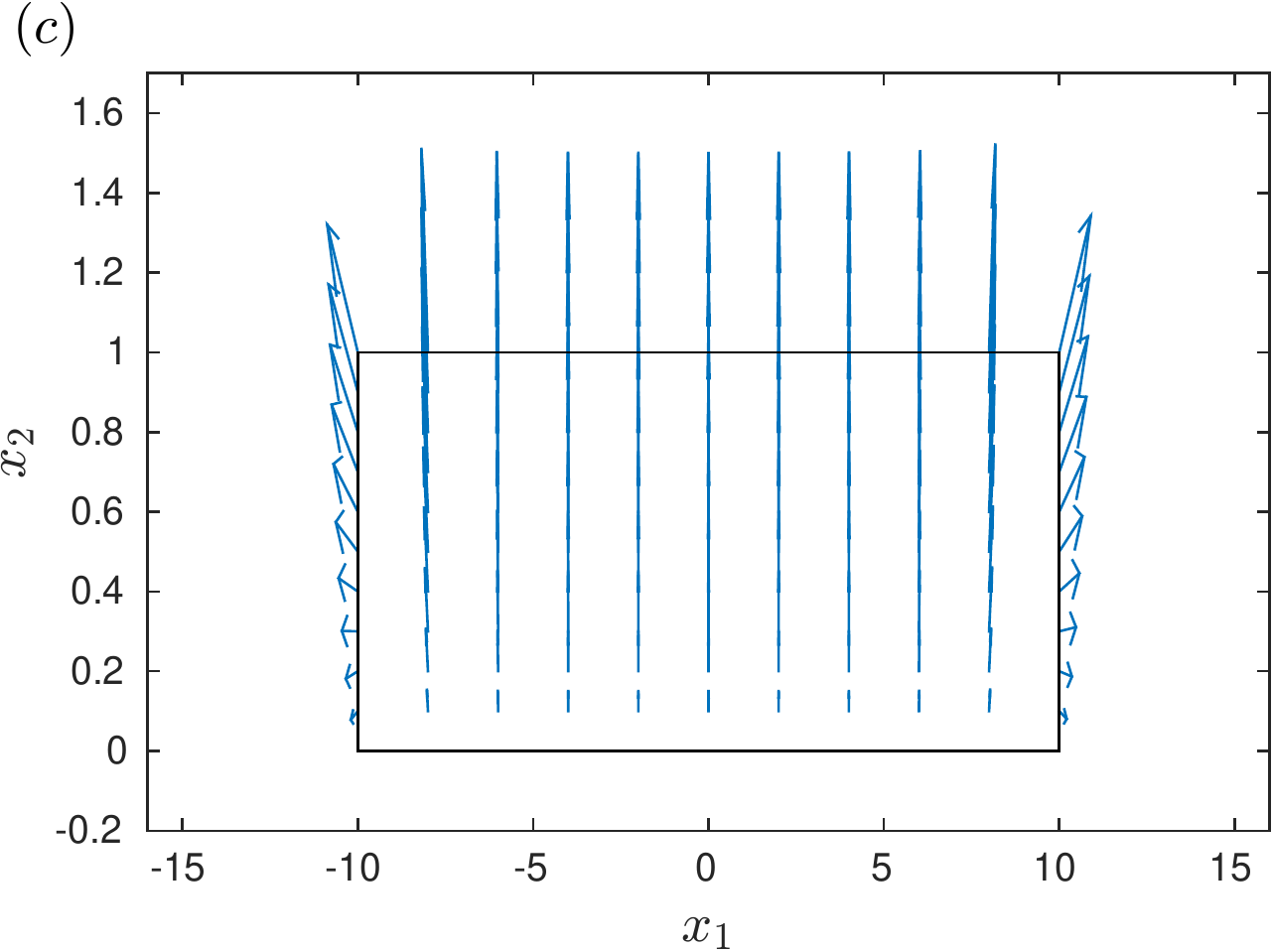}
\includegraphics[width=0.4\textwidth]{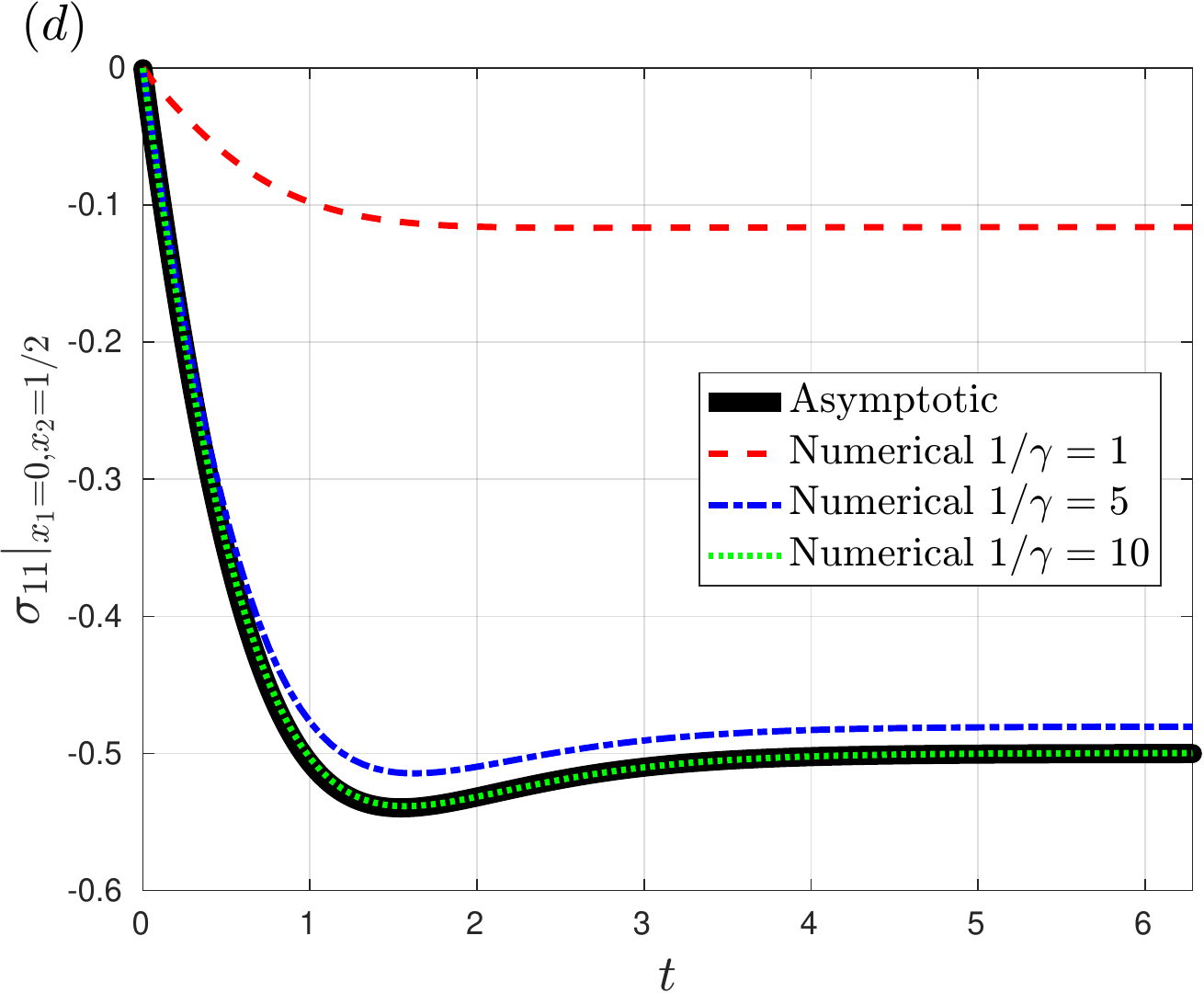}
\caption{Panels (a)-(c) show numerical solutions for the deformation
  field $(u_1,u_2)$ to the macroscale equations
  (\ref{macro_start})-(\ref{end_macro_bcs}) for different values of
  $1/\gamma=1,5,10$ respectively. Convergence to a one-dimensional
  deformation is clearly observed as $\gamma
  \rightarrow 0^+$. Panel (d) shows a comparison between
  $\sigma_{11}|_{x_1=0,x_2=1/2} (t)$ from the numerical simulations
  and the asymptotic analysis, i.e.~(\ref{macro_sol_end}).}
\label{macro_def_examples}
\end{figure}

We note that the one-dimensional solution we have obtained only
applies throughout the bu
lk of the electrode, and does not satisfy the
stress-free conditions imposed on $\Gamma_{2,4}$. This apparent
discrepancy can be resolved by rescaling spatial variables
appropriately and studying the solution in boundary layers local to
the lateral edges of the electrode. In the interests of brevity we do
not present the details of this analysis here. Instead, to provide
evidence that the one-dimensional solution is indeed correct
in the bulk, we solve the problem
(\ref{macro_start})-(\ref{end_macro_bcs}) numerically and verify
agreement with the analytical solution (details of the numerical
  approach are presented in \S\ref{sec:numer}). A selection of
numerical solutions with increasing aspect ratio (decreasing $\gamma$)
is shown in Figure \ref{macro_def_examples} --- as one can see,
the agreement throughout the bulk improves as $\gamma \to 0^+$ and the two solutions are almost indistinguishable for $1/\gamma=10$.

\section{\label{time_passes}The microscale problem}
We now revisit the microscale problem, rescaling the governing
  equations about a generic individual electrode particle at
$(x_1,x_2)=(x_{1,m},x_{2,n})$ somewhere in the bulk of the electrode
by writing
\begin{equation}
x_2-x_{2,n} = \lambda X_2, \qquad x_1-x_{1,m} = \lambda X_1, \qquad u_1 = v_{1,m,n} + \lambda U_1, \qquad u_2 = v_{2,m,n} + \lambda U_2.
\end{equation}
We aim to solve (\ref{nd_start})-(\ref{smell_my_cheese}) for the
microscale variables $X_i$, $U_i$ on a single periodic unit cell
$-1/2<X_1<1/2$, $-1/2<X_2<1/2$.  Boundary conditions on the microscale
cell problem need to be chosen in order that the micro-solution is
compatible with the one-dimensional solution to the macroscale problem
(\ref{deaf}). Since there is no macroscopic shear, we require
$\sigma_{12} = 0$ on $X_1 = \pm 1/2$ and $X_2 = \pm 1/2$. In
  addition, since there are no lateral macroscopic displacements, we
require $U_1 = 0$ on $X_2 = \pm1/2$.  On the other hand,
averaging the 22 component of the microscopic strain over the
  unit cell, which yields the net relative displacement of the
  horizontal boundaries of the cell, should be also equal to the
  macroscopic strain, so that
\begin{equation*}
\int_{-1/2}^{1/2} \frac{\partial U_2}{\partial X_2}\, dX_2 = 
U_2\big|_{X_2 = 1/2} - U_2\big|_{X_2 = -1/2}  = \epsilon^{\mathrm{eff}}_{22} = \frac{\partial u_2^{\mathrm{eff}}}{\partial x_2}= 2 \ell(t).
\end{equation*}
Since we have not derived the effective coefficients of the macroscale
problem, and therefore cannot accurately  determine the macroscale
strain $\epsilon^{\mathrm{eff}}_{22}$,
we do not impose $\ell(t)$ directly but calculate it self-consistently using the
condition that there is no normal macroscale stress in the vertical
direction ($\sigma_{22}^{\mathrm{eff}}=0$), which implies
\be 
\int_{-1/2}^{1/2} \sigma_{22} \, \d X_1=0 \quad \mbox{ on } \quad X_2 = \pm 1/2.
\ee
Finally we have the swelling condition that $U_i = X_i g(t)$ on the
particle $\Gamma_5^{m,n}$.

\begin{figure}          \centering
\includegraphics[width=0.4\textwidth]{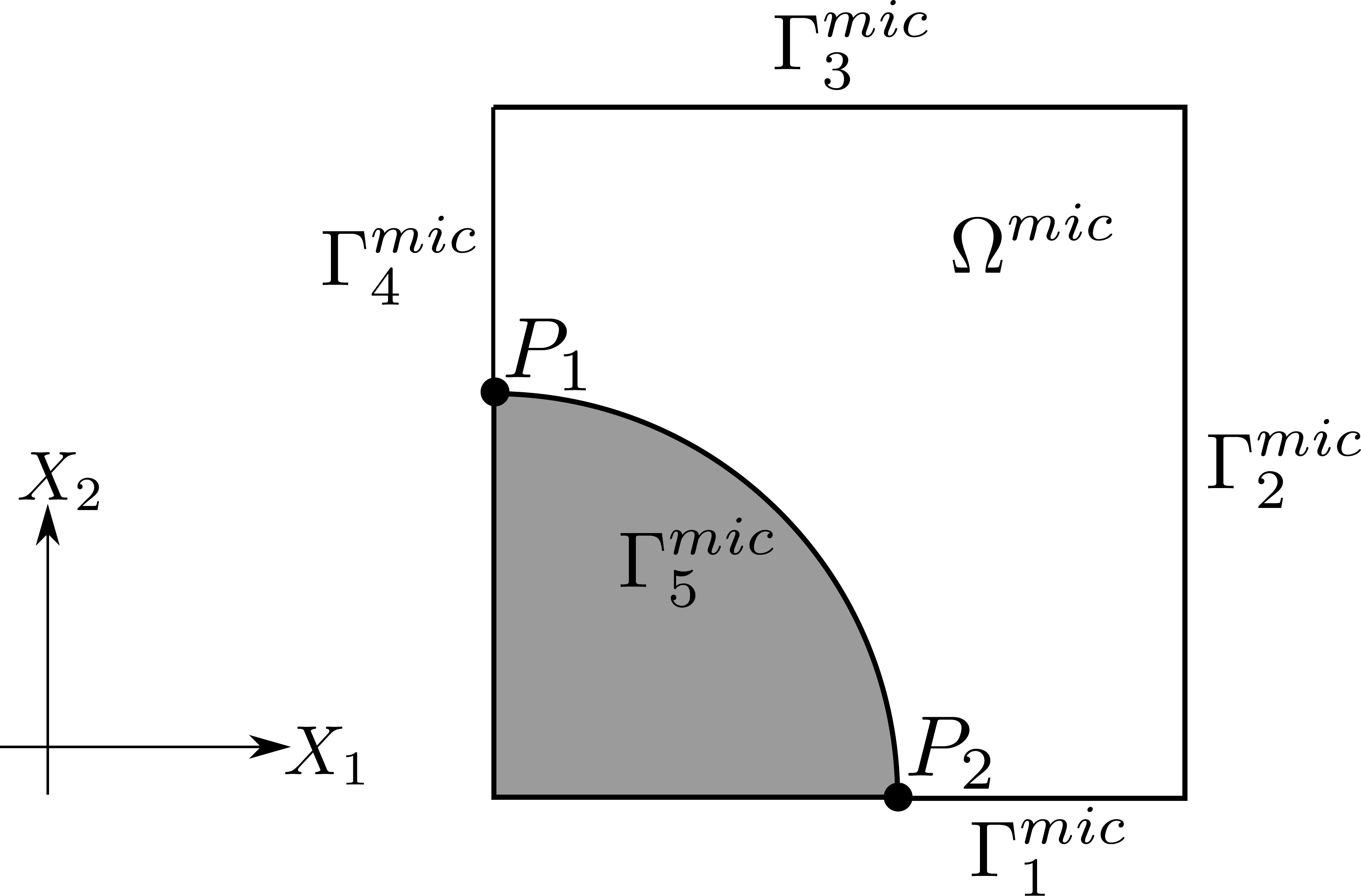}
\caption{Schematic of the geometry of the microscale problem.}
\label{micro_scheme}
\end{figure}

The microscale problem we have derived is symmetric in both $X_1$ and
$X_2$, so that we may solve it more efficiently by considering 
 the
quarter-cell domain, depicted in Figure \ref{micro_scheme}, with
boundary segments
\begin{eqnarray}
& \label{quarterstart} \Gamma_1^{\mathrm{mic}} =  \{X_1,X_2 \, |\, r_0<X_1<1/2,X_2=0\}, \qquad \Gamma_2^{\mathrm{mic}} =  \{X_1,X_2 \, |\, X_1=1/2,0<X_2<1/2\},\\
& \Gamma_3^{\mathrm{mic}} =  \{X_1,X_2 \, |\, 0<X_1<1/2, X_2=1/2\}, \quad \Gamma_4^{\mathrm{mic}} =  \{X_1,X_2 \, |\, X_1=0,r_0<X_2<1/2\},\\
& \label{quarterend} \Gamma_5^{\mathrm{mic}}  = \{X_1,X_2 \, | \,
  X_1>0,X_2>0,X_1^2+X_2^2=r_0^2\}. & 
\end{eqnarray}
The boundary and symmetry conditions are then

\vspace{-1.5em}
\noindent\begin{tabularx}{\textwidth}{@{}XXX@{}}
\begin{equation} \mbox{on} \, \, \Gamma_1^{\mathrm{mic}}:
  \nonumber \end{equation} & \begin{equation}
  U_2=0 \label{cob_salad} \end{equation} & \begin{equation}
  \sigma_{12} = 0, \end{equation} 
\end{tabularx}

\vspace{-4em}
\noindent\begin{tabularx}{\textwidth}{@{}XXX@{}} 
\begin{equation} \mbox{on} \, \, \Gamma_2^{\mathrm{mic}}:
  \nonumber \end{equation} & \begin{equation} U_{1}=0  \end{equation}
& \begin{equation} \sigma_{12} = 0, \end{equation} 
\end{tabularx}

\vspace{-4em}
\noindent\begin{tabularx}{\textwidth}{@{}XXX@{}}
\begin{equation} \mbox{on} \, \, \Gamma_3^{\mathrm{mic}}:
  \nonumber \end{equation} & \begin{equation}
  U_{2}=\ell(t)  \end{equation} & \begin{equation} \sigma_{12} =
  0, \end{equation} 
\end{tabularx}

\vspace{-4em}
\noindent\begin{tabularx}{\textwidth}{@{}XXX@{}}
\begin{equation} \mbox{on} \, \, \Gamma_4^{\mathrm{mic}}:
  \nonumber \end{equation} & \begin{equation} U_{1}=0  \end{equation}
& \begin{equation} \sigma_{12} = 0, \end{equation} 
\end{tabularx}

\vspace{-4em}
\noindent\begin{tabularx}{\textwidth}{@{}XXX@{}}
\begin{equation} \mbox{on} \, \, \Gamma_5^{\mathrm{mic}}:
  \nonumber \end{equation} & \begin{equation} U_{1}= X_1
  g(t)  \end{equation} & \begin{equation} \label{beef} U_2 = X_2
  g(t), \end{equation} 
\end{tabularx}
where $\ell(t)$ is determined by the requirement that \be
\label{constraint} \int_{\Gamma_3^{\mathrm{mic}}} \sigma_{22}\, \d
X_1=0.  \ee Finally, we note that the condition
(\ref{smell_my_cheese}) is automatically satisfied due to symmetry and
also that there is no net torque on the particle, so that our earlier
decision not to allow for rotations is justified.

Owing to the geometry, solution of the microscale problem must be
obtained using numerical techniques. We employ the finite element
method, and implement this approach using the open source software
{\tt FreeFEM++} \cite{freefem}. The problem has two non-standard
features, namely: (I) the constitutive equations involve time
derivatives of the stress- and strain-fields, and; (II) the problem is
subject to an integral constraint, namely (\ref{constraint}). The
first of these difficulties is tackled by discretizing the governing
equations in time using an explicit approximation; this leads to
an elliptic boundary-value problem for the state variables at
each time step that are of similar form to those for an elastic medium
--- information from the previous time step is retained in the form of
source terms. The second difficulty is circumvented by embedding a
code that solves the problem for a specified value of $\ell(t)$ (a
tractable problem) within a root-finding loop that determines the
value of $\ell(t)$ at the new time step that leads to a stress field
satisfying (\ref{constraint}). Since the variation of the average
normal load on the top surface is a linear function of $\ell(t)$, only
two `test' problems need to be solved per time step in order to find
the value of $\ell(t)$ satisfying (\ref{constraint}). We now detail the
method of numerical solution to this microscale problem.

\subsection{\label{sec:numer}Numerical approach}
Before implementation in {\tt FreeFem++}, the system of equations must
be written in a suitable variational form. Given the non-standard
nature of our problem, we briefly outline the procedure for deriving
this variational formulation. First, we take the partial derivative of
the force balance equation (\ref{nd_start}) with respect to time and
then eliminate $\sigma_{ij}$ in favor of $\epsilon_{ij}$, $E$ and $S$
using equations (\ref{ee3})-(\ref{nd_eq_end}). On doing so, and
denoting a partial derivative in time with a dot, we arrive at
\begin{equation} \begin{split}
\frac{\partial}{\partial x_j} \left( G_2 G_\tau \dot{\epsilon}_{ij} + \epsilon_{ij} \right. \qquad \qquad \qquad \qquad \qquad \qquad \qquad \qquad \qquad \qquad \qquad \qquad \\ \left. + \delta_{ij} \left[ (K_2 K_\tau-G_2 G_\tau) \dot{E} - K_2 K_\tau \dot{\beta}_{\mathrm{abs}} + (K_1-1)E - K_1 \beta_{\mathrm{abs}} \right]\right) =  0.
\end{split} \end{equation} This equation is then discretized
explicitly in time using a forward Euler approximation, multiplied by
a test function $v_i \in C^{\infty}_0(\Omega)$ and integrated by
  parts (in the usual way) to obtain the variational form
\begin{eqnarray}
\int_\Gamma a_1 \epsilon_{ij} (u^{(n+1)}) v_j n_i + a_2 \epsilon_{ij} (u^{(n)}) v_j n_i + a_3 E(u^{(n+1)}) v_i n_i + a_4 E(u^{(n)}) v_i n_i \, dS \nonumber \\
-\int_\Omega a_1 \epsilon_{ij} (u^{(n+1)}) \epsilon_{ij} (v) + a_2 \epsilon_{ij} (u^{(n)})\epsilon_{ij} (v) + 2 a_3 E(u^{(n+1)}) E(v) \label{mess} \\ 
+ 2 a_4 E (u^{(n)}) E(v) - G_\tau \frac{\partial S^{(n)}}{\partial x_i} v_i \,d\Omega. \nonumber
\end{eqnarray}
Here the superscripts denote the respective time levels, $\Delta t$ is the (small) time step and the coefficients ($a_1$ to $a_4$) are defined by the relations
\begin{eqnarray}
a_1 = G_2 G_\tau,\\
a_2 = \Delta t - G_2 G_\tau,\\
a_3 = K_2 K_\tau - G_2 G_\tau,\\
a_4 = \Delta t(K_1 - 1) -(K_2 K_\tau - G_2 G_\tau).
\end{eqnarray}
On using the boundary conditions (\ref{cob_salad})-(\ref{beef}) to
simplify the boundary term, equation (\ref{mess}) can be viewed as a
problem for the displacement (and in turn the strain) field for a
given value of the volumetric stress at the previous time step
$S^{(n)}$. An equation for the evolution of $S^{(n)}$ is obtained by
discretizing (\ref{nd_eq_end}) in time (again using the forward
  Euler method) to give
\begin{equation} \begin{split} \label{queen_elizabeth}
S^{(n+1)} = \left( 1 - \frac{\Delta t}{K_\tau} \right) S^{(n)} + K_2 E^{(n+1)} \qquad \qquad \qquad \qquad \qquad \qquad \qquad \qquad  \\ + \left( \frac{K_1 \Delta t}{K_\tau} - K_2 \right) E^{(n)} - K_2 \beta_{\mathrm{abs}}^{(n+1)} + \left( K_2 - \frac{K_1 \Delta t}{K_\tau} \right) \beta_{\mathrm{abs}}^{(n)}.
\end{split} \end{equation} A code that evolves (\ref{mess}) and
(\ref{queen_elizabeth}) may then be implemented in {\tt FreeFem++} ---
a pseudo-code is included in algorithm \ref{alg1}.

\begin{algorithm}
\begin{algorithmic}
\STATE $t=0$
\STATE $n=0$
\STATE Define the initial conditions: $u_i^{(0)}=\epsilon_{ij}^{(0)}=E^{(0)}=S^{(0)}=L^{(0)}=0$
\REPEAT
\STATE Advance time: $t=t+\Delta t$
\STATE Define the first `test' value of $\ell^{(n+1)}$: $\ell_1' = \ell^{(n)} - 0.1$\\ 
\STATE Define the second `test' value of $\ell^{(n+1)}$: $\ell_2' = \ell^{(n)} + 0.1$\\
\STATE Solve (\ref{mess}) with $\ell^{(n+1)} = \ell_1'$\\
\STATE Evaluate the average normal load on $\Gamma_{\mathrm{mic}}^3$ for the first `test' value: $F_1 = \int_{\Gamma_3^{\mathrm{mic}}} \sigma^{(n+1)}_{yy} dS$\\
\STATE Solve (\ref{mess}) with $\ell^{(k+1)} = \ell_2'$\\
\STATE Evaluate the average normal load on $\Gamma_{\mathrm{mic}}^3$ for the second `test' value: $F_2 = \int_{\Gamma_3^{\mathrm{mic}}} \sigma^{(n+1)}_{yy} dS$\\
\STATE Compute the value of $\ell^{(n+1)}$ corresponding to zero load on $\Gamma_{\mathrm{mic}}^3$: $\ell^{(n+1)} = \ell_2' - F_2  ((\ell_2' - \ell_1')/(F_2 - F_1))$\\
\STATE Solve (\ref{mess}) for $u_i^{(n+1)}$, $\epsilon_{ij}^{(n+1)}$ and $E^{(n+1)}$ using $\ell^{(n+1)}$ \\
\STATE Compute $S^{(n+1)}$ using (\ref{queen_elizabeth})  \\
\STATE Advance the index: $n=n+1$
\UNTIL $t>T_{end}$
\end{algorithmic}
\caption{The pseudo-code for the evolution on the microscopic scale.}
\label{alg1}
\end{algorithm}

All the usual diagnostic checks were carried out to ensure that the
method was performing as expected. In order to ensure errors were
acceptably small, the number of elements (and the size of the time
step) were increased (decreased) until no appreciable changes were
observed in the solutions. Typically, we found that using $P2$
elements, meshing the domain with approximately $100 \times 100$
elements, and taking the time step $\Delta t = 10^{-3}$ gave solutions
which were converged to 4 significant digits of accuracy.

\subsection{\label{poly_swell}Deformation driven by electrolyte absorption}
In this section we consider forcing deformation in the microscale
problem by allowing the polymer binder to absorb electrolyte, i.e.~via
the function $\beta_{\mathrm{abs}}$ in (\ref{nd_eq_end}) --- the
volumetric change of the electrode particle is taken to be zero. In
line with the estimates given in \S\ref{scalings} we take
\begin{eqnarray}
\label{america1} \beta_{\mathrm{abs}} = \frac{1}{2} \tanh(t), \qquad g(t) = 0 \qquad r_0 = \frac{1}{2}, \qquad G_{\tau} = 0.02,\\
\label{america2} G_2 = 3, \qquad K_\tau = 0.02, \qquad K_1 = 1, \qquad K_2 = 3,
\end{eqnarray} 
so that the absorption of the electrolyte causes the volume of the polymer (corresponding to zero stress) to increase by 50\% for large time.

\begin{figure}          \centering
\includegraphics[width=0.4\textwidth]{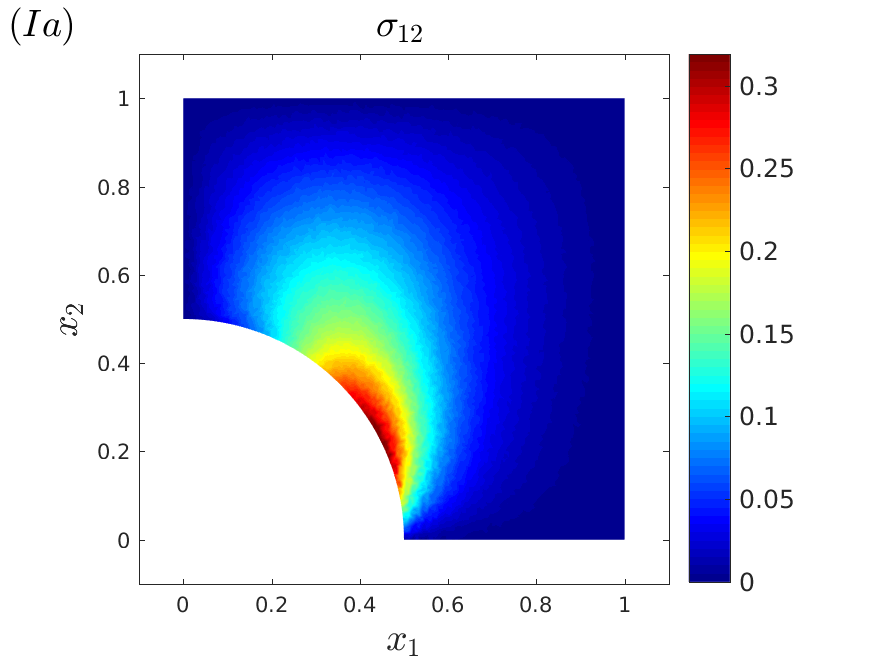}
\includegraphics[width=0.4\textwidth]{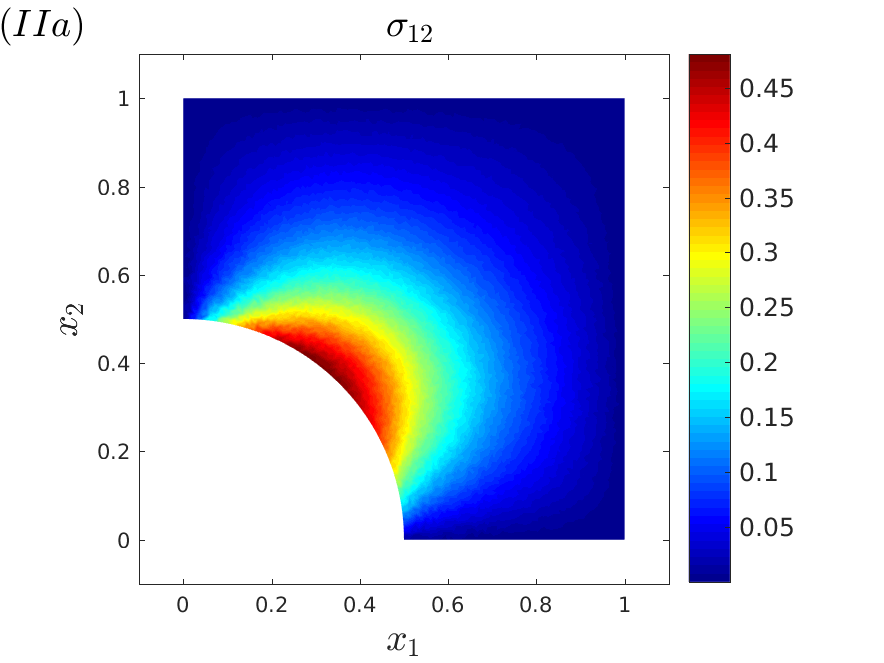}

\includegraphics[width=0.4\textwidth]{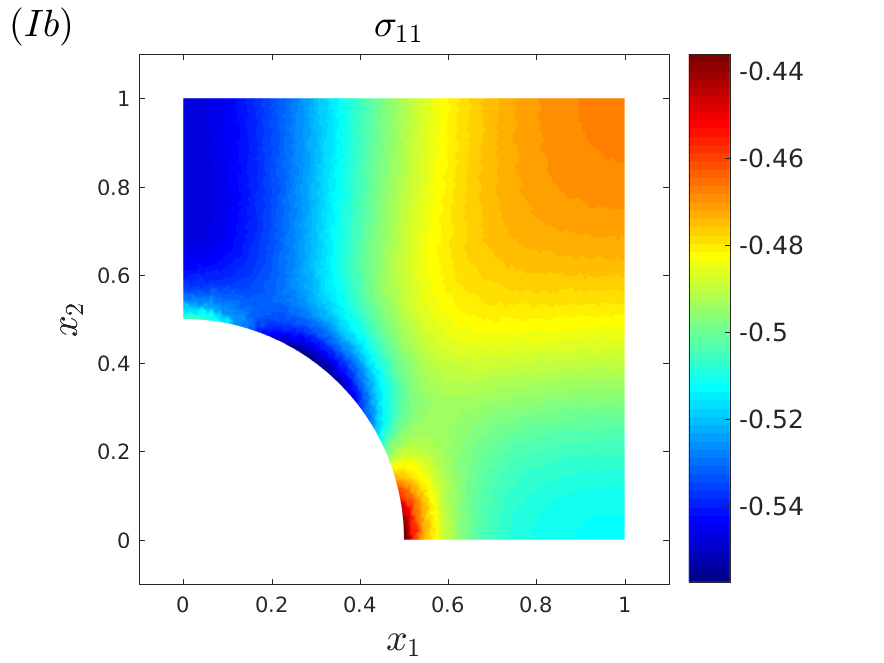}
\includegraphics[width=0.4\textwidth]{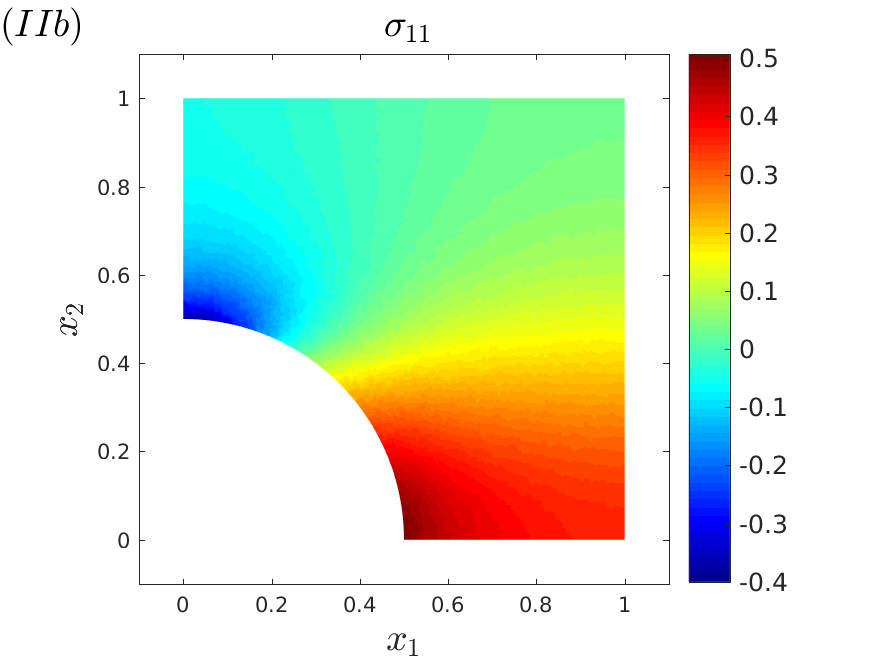}

\includegraphics[width=0.4\textwidth]{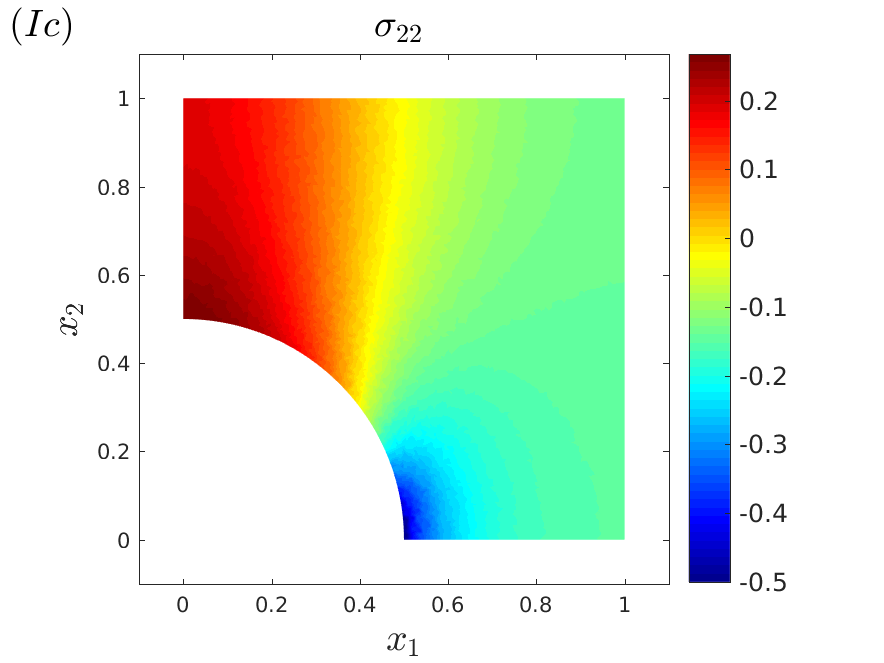}
\includegraphics[width=0.4\textwidth]{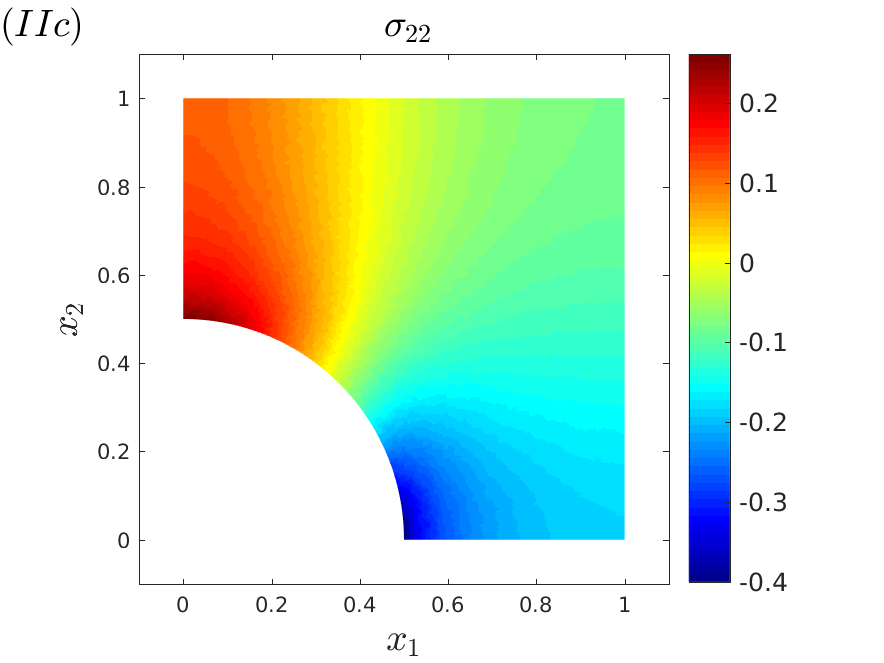}

\includegraphics[width=0.4\textwidth]{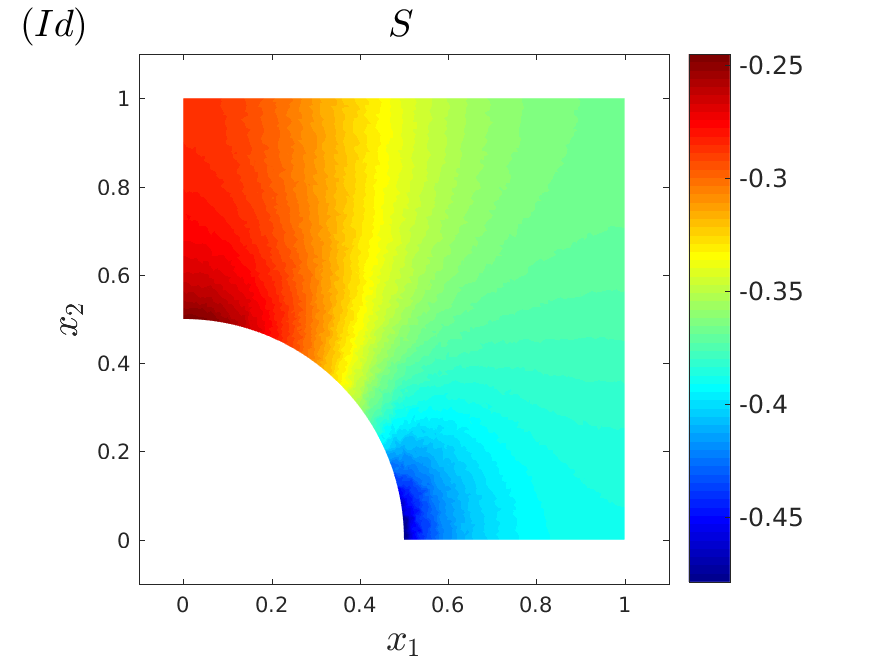}
\includegraphics[width=0.4\textwidth]{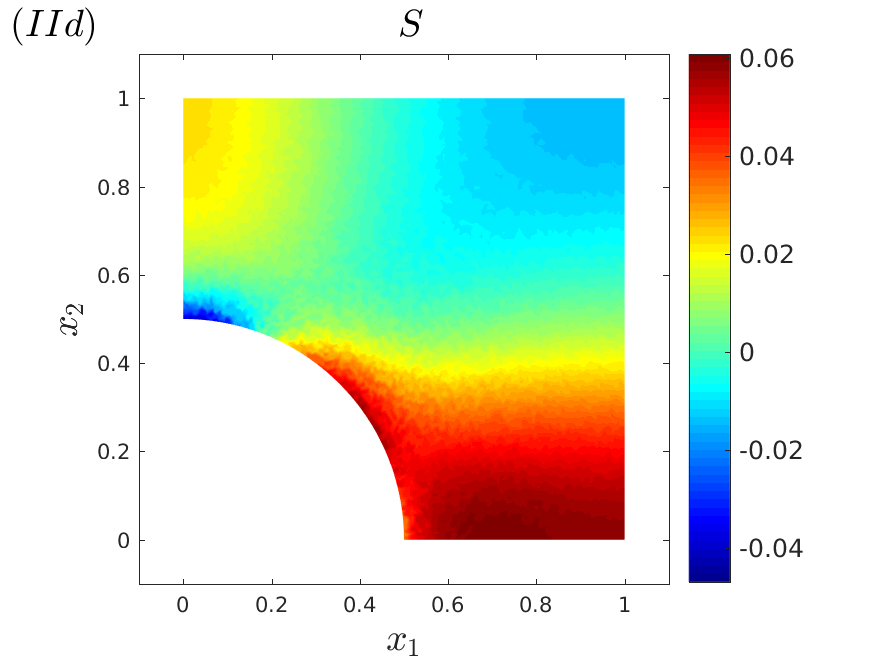}

\caption{Column (I) shows the different components of the stress field
  at the end of the simulation defined by
  (\ref{america1})-(\ref{america2}), i.e.~after the binder has
  absorbed electrolyte and swelled. Here, we see that
  $\sigma_{22}|_{P_1}>0$ (binder in tension) whereas
  $\sigma_{11}|_{P_2}<0$ (binder in compression). Column (II)
  shows analogous plots at $t=\pi$ for the simulation defined by
  (\ref{ohyeah1})-(\ref{ohyeah2}), i.e.~for deformations driven by
  cycling in a cathode at the peak of discharge. Here, we see that
  both $\sigma_{22}|_{P_1},\sigma_{11}|_{P_2}>0$ (binder
  in tension).}
\label{elec_abs_fig}
\end{figure}

\begin{figure}          \centering
\includegraphics[width=0.35\textwidth]{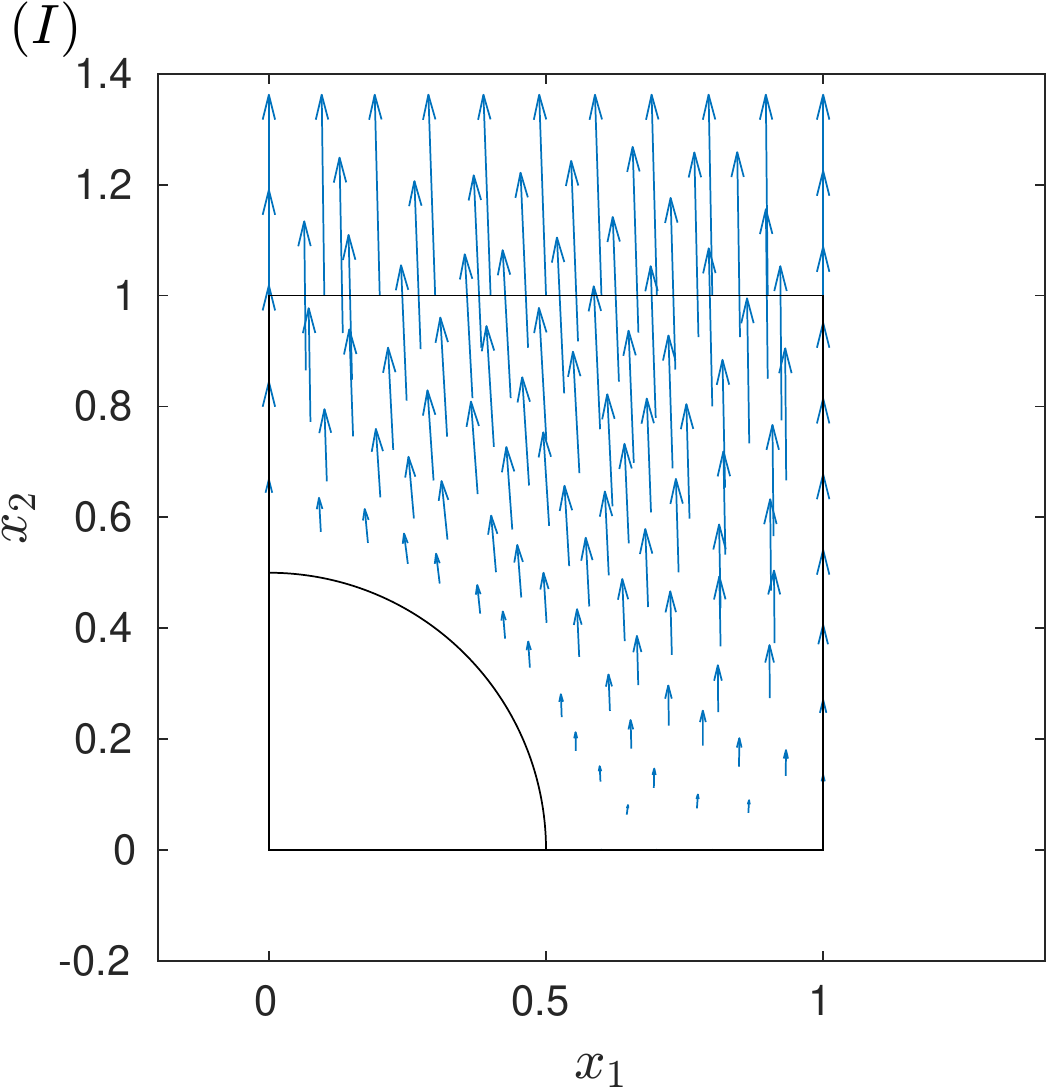} \hspace{5em}
\includegraphics[width=0.35\textwidth]{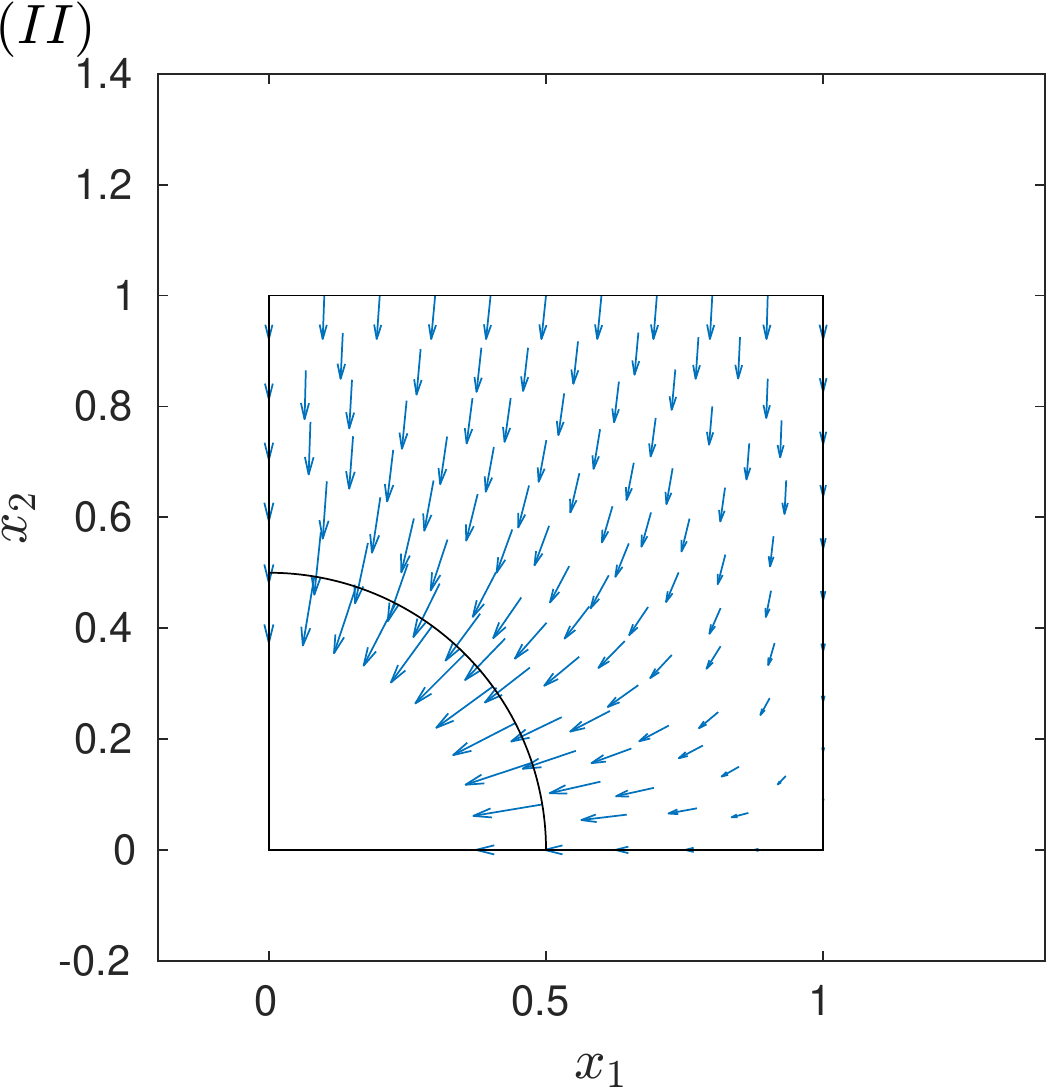}
\caption{Panel (I) shows the deformation field at the end of the
  simulation defined by (\ref{america1})-(\ref{america2}), i.e.~after
  the binder has absorbed electrolyte and swelled. Panel (II) shows
  the deformation field at $t=\pi$ for the simulation defined by
  (\ref{ohyeah1})-(\ref{ohyeah2}), i.e.~for deformations driven by
  cycling in a cathode at the peak of discharge. Arrow lengths have
  been normalized in each plot individually.}
\label{deforms}
\end{figure}

\begin{figure}          \centering
\includegraphics[width=0.45\textwidth]{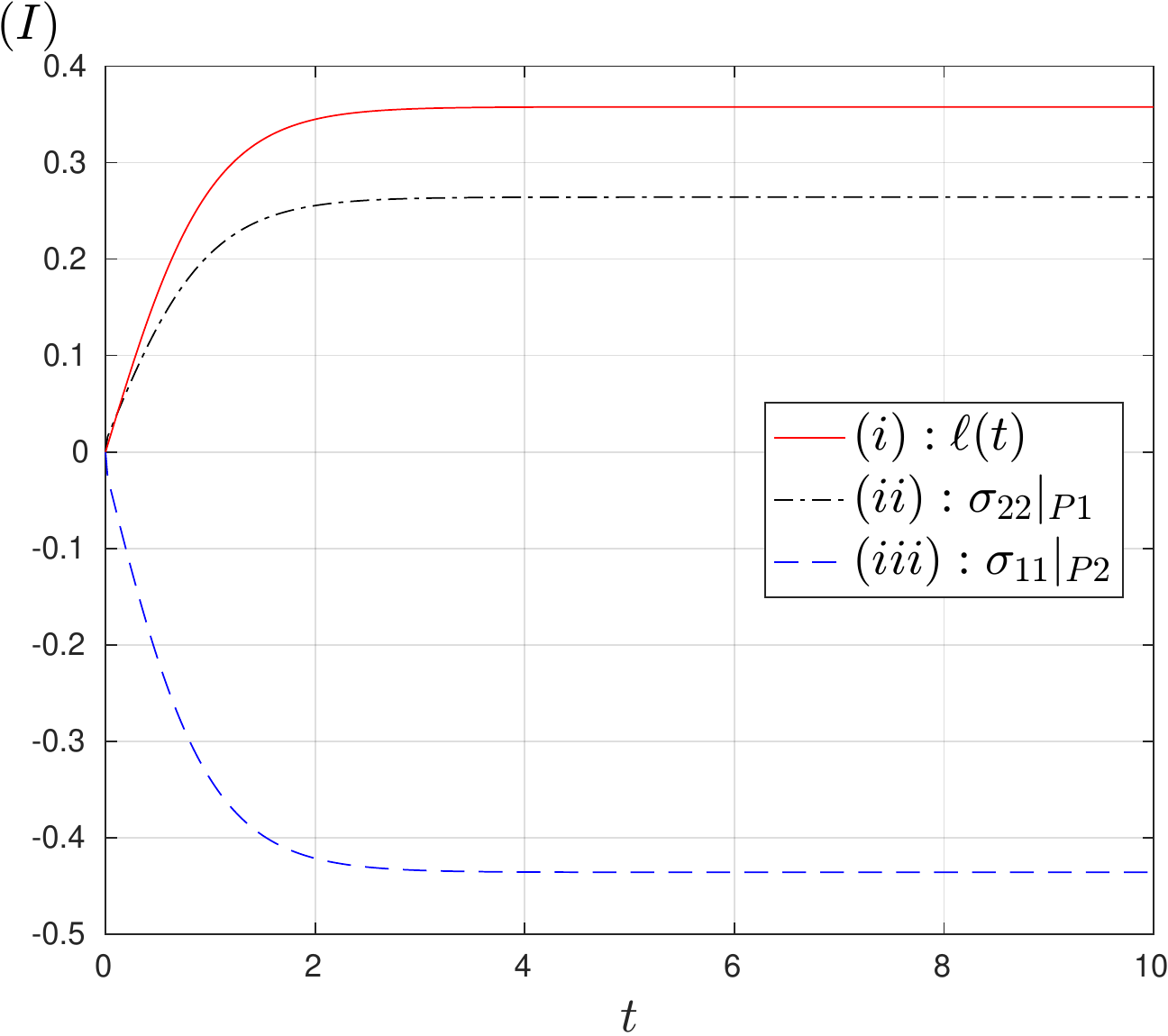}
\includegraphics[width=0.48\textwidth]{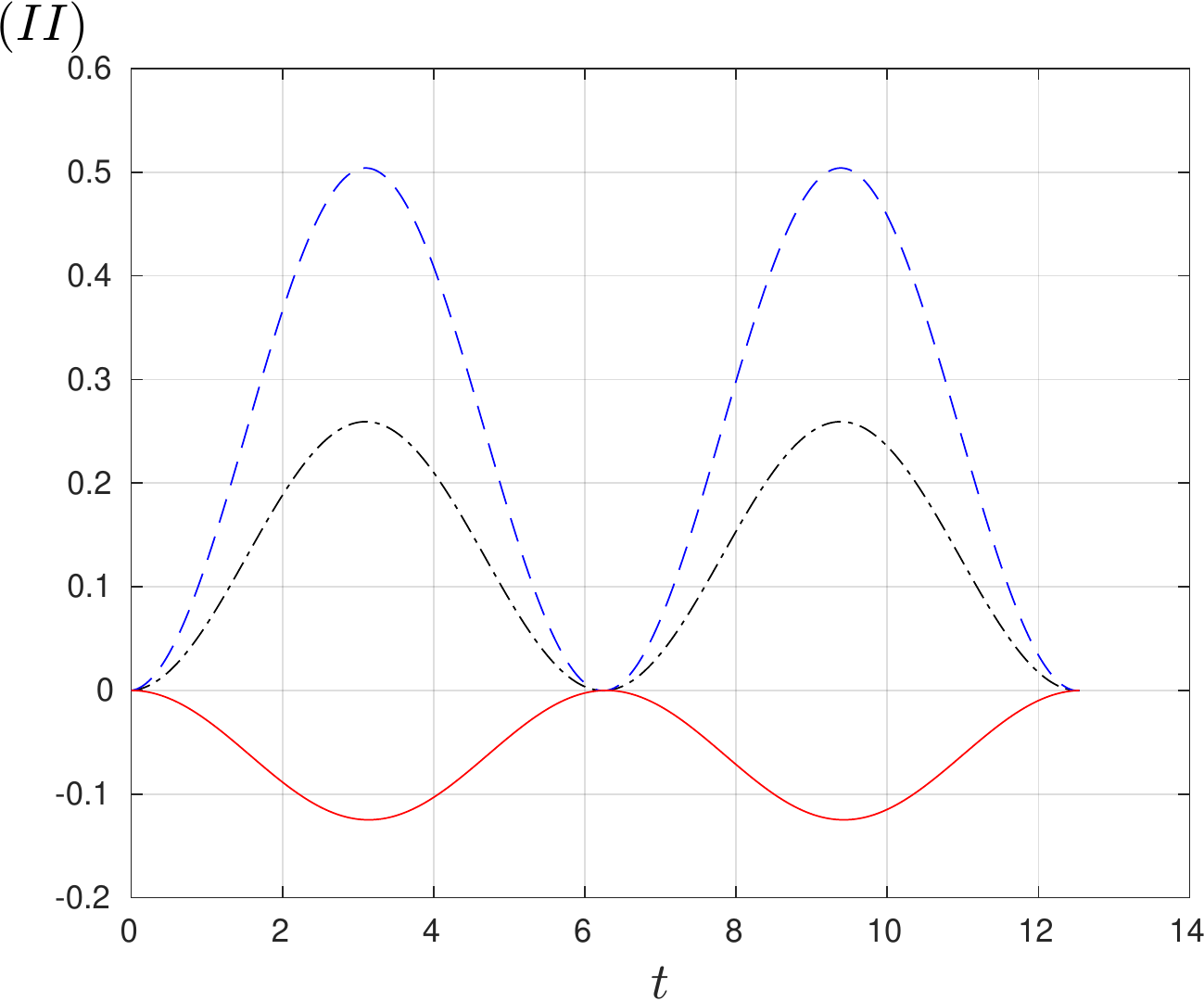}
\caption{The evolution of: (i) $\ell(t)$, the vertical displacement on
  the top of the unit cell $\Gamma_3^{\mathrm{mic}}$; (ii)
  $\sigma_{22}|_{P_1}$, the normal stress at the point $P_1$,
  and; (iii) $\sigma_{11}|_{P_2}$, the normal stress at the
  point $P_2$. Panel (I) shows the results for the simulation defined
  by (\ref{america1})-(\ref{america2}), i.e.~for deformations driven
  by electrolyte absorption, whereas panel (II) shows the results for
  the simulation defined by (\ref{ohyeah1})-(\ref{ohyeah2}), i.e.~for
  deformations driven by cycling.}
\label{curves}
\end{figure}

The stress fields generated at the end of the simulation are shown in
Figure \ref{elec_abs_fig}, the deformation field at the end of the
simulation (at $t=T_{end}=10$) is shown in Figure \ref{deforms} whilst the evolution of
both $\ell(t)$ and the normal stresses at the points $P_1$ and $P_2$ (as
defined in Figure \ref{micro_scheme}) are shown in Figure
\ref{curves}. We see that after swelling, the normal stress at the top
(and bottom) of the electrode particle is tensile whereas at the
right-hand (and left-hand) side it is compressive. This indicates that
deformations induced by electrolyte absorption are likely to cause
preferential delamination of the polymer binder from the top (and
bottom) of the particle surfaces. This result can be understood
intuitively by considering a growing solid medium confined within two
rigid vertical walls. If the material were to expand without
constraint it would dilate uniformly, however because it is confined
in the horizontal direction, it has no choice but decrease its
internal stresses by flowing/deforming predominantly in the vertical
direction --- see panel (a) of Figure \ref{deforms}. If this material
were also to contain a circular rigid inclusion ({e.g.} an electrode
particle), then the description above is still appropriate and the
growing material would therefore try to pull away from the circular
inclusion at its top and bottom surface as it flows/deforms.

In Figure \ref{curves} we see that for the parameter choices given in
(\ref{america1})-(\ref{america2}) the increase in stress at the point
P1 (and decrease at point P2) is monotonic in time and saturates to a
constant. The constant to which these quantities tend is controlled
by: (i) the long time behavior of the growth function
$\beta_{\mathrm{abs}}(t)$, and; (ii) the long term moduli of the
polymer, $K_1$ --- corresponding to $G_1^*$ and $K_1^*$ in dimensional
quantities. If one were designing a binder specifically to minimize
the amount of damage caused by polymer swelling one could therefore
try to minimize the degree of absorption and/or the long term moduli.

In addition to the results shown in Figures \ref{elec_abs_fig},
\ref{deforms} and \ref{curves}, other simulations with different
material constants were also carried out. It is possible to
qualitatively change the monotonic behavior seen in Figure
\ref{curves} if the time scales for absorption and viscoelastic
relaxation become comparable, i.e.~if $G_\tau$ or $K_\tau=O(1)$. In
such regimes, transient behaviour may occur with the magnitude
of the normal stresses at the point $P_1$ and $P_2$ increasing
to large values before relaxing back down to those predicted by their
long term moduli. In the context of mitigating mechanical damage this
is an undesirable situation. Thus, one further practical
recommendation is to ensure that the viscoelastic relaxation
timescales remain much shorter than those for electrolyte absorption
thereby taking advantage of the tendency of viscoelastic materials to
flow in order to minimize their internal stresses. As far as the
authors are aware, most polymers used for binder do take a long time
to absorb electrolyte (at least several hours) and therefore it seems
unlikely that this viscoelastic relaxation would ever be observed in
practice.

\subsection{\label{AM_swell}Deformation driven by electrode particle swelling}
Here, we investigate the evolution as the electrode particle changes
its volume. We study a situation analogous to the images shown in
Figure \ref{example_slices}, i.e.~a cathode in which the
particles are at their largest at the time of manufacture; as the
device is discharged the cathode particles shrink and then at some
later time, as the device is re-charged, the cathodic active material
grows. We base our initial simulation on electrode particles composed
of NMC and, in line with the discussion in \S\ref{scalings}, we
take
\begin{eqnarray}
\label{ohyeah1} \beta_{\mathrm{abs}} = 0, \qquad g(t) = \frac{\cos(t)-1}{10}, \qquad r_0=\frac{1}{2}, \qquad G_{\tau} = 0.02,\\
\label{ohyeah2} G_2 = 3, \qquad K_\tau = 0.02, \qquad K_1 = 1, \qquad K_2 = 3.
\end{eqnarray}
The components of the stress field at the end of the simulation (at $t=T_{end}=4\pi$),
i.e.~after two complete cycles, are shown in the right column of
Figure \ref{elec_abs_fig}, the corresponding deformation field is
shown in Figure \ref{deforms} whilst the evolution of the normal
stress at the points $P_1$ and $P_2$ on the surface of the electrode
particle is shown in Figure \ref{curves}. In contrast to the
deformations driven by electrolyte absorption, these electrode
particle-induced deformations cause tension in the binder at all
positions on the particle surface. Further, the normal stress
(directed outward) at the right-hand (and left-hand) sides of the
electrode particle are larger than those at the top (and bottom),
indicating that volumetric changes in the electrode particles are
likely to drive delamination of the binder at the left and right edges
of the electrode particle more strongly. The reasons for this are
straightforward to understand: throughout the cycle the particles are
smaller than their original size and therefore the surrounding polymer
material is in tension. The size of this tension is larger at the
left- and right-hand sides, because the adjacent sides of the unit cell
are fixed, whereas the top (and bottom) of the unit cell can deform
thereby reducing the tension.

Since these simulation were based on electrochemical cycling taking
place on the scale of tens of hours (comparable to the timescale for
electrolyte absorption), the viscoelastic relaxation is not clearly
observed. The size of the stresses are therefore primarily controlled
by the material parameter $K_1$ --- or $K_1^*$ and $G_1^*$ in
dimensional form --- and the size of the volumetric changes of the
particle $g(t)$.

\begin{figure}          \centering
\includegraphics[width=0.6\textwidth]{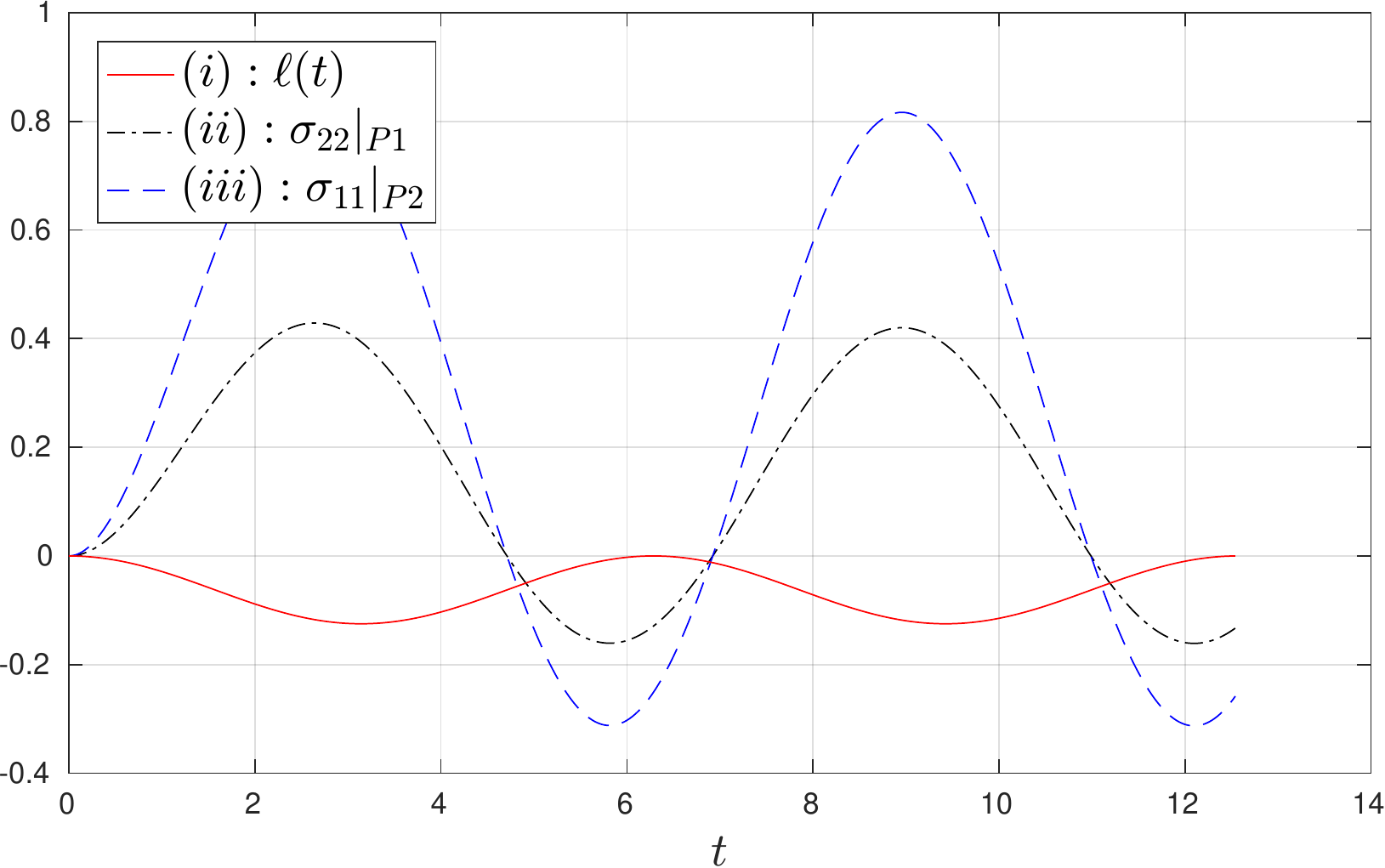}
\caption{The evolution of: (i) $\ell(t)$, the vertical displacement
  of the top of the unit cell $\Gamma_3^{\mathrm{mic}}$; (ii)
  $\sigma_{22}|_{P_1}$, the normal stress at the point $P_1$,
  and; (iii) $\sigma_{11}|_{P_2}$, the normal stress at the
  point $P_2$. The result shown is for a cathode under aggressive
  cycling conditions as discussed in the penultimate paragraph of
  \S\ref{AM_swell}.}
\label{other_curves}
\end{figure}

It is relevant to also consider the evolution when the cathode is
cycled more aggressively. To do so we keep all parameters identical to
those shown in (\ref{ohyeah1})-(\ref{ohyeah2}), but time is rescaled
so that the relaxation timescales $G_\tau = K_\tau = 1$. Thus, the
cycling rate has been increased by a factor of 50 so that a cycle now
takes only several minutes (rather than hours). The evolution of the
normal stresses at $P_1$ and $P_2$ as well as $\ell(t)$ are shown in
Figure \ref{other_curves}. The viscoelastic relaxation is now
apparent --- the stresses are no longer close to zero at the end of each
cycle. When the particle has returned to its original size, the normal
stresses have become negative so that the surrounding binder is under
compression; a beneficial configuration in the context of mitigating
delamination. This is observed because the binder is in tension
mid-cycle (the particle is smaller than originally).
Thus, when the particle returns to its original size, the
binder has undergone creep so that its zero stress state corresponds
to a configuration with a smaller embedded electrode particle. This
creeping effect is stronger at $P_2$ than at $P_1$, because the
vertical edges of the unit cell are fixed whereas the top (and bottom)
can move thereby causing the binder at the top (and bottom) to be
under a decreased load leading to reduced creep.

\begin{figure}          \centering
\includegraphics[width=0.48\textwidth]{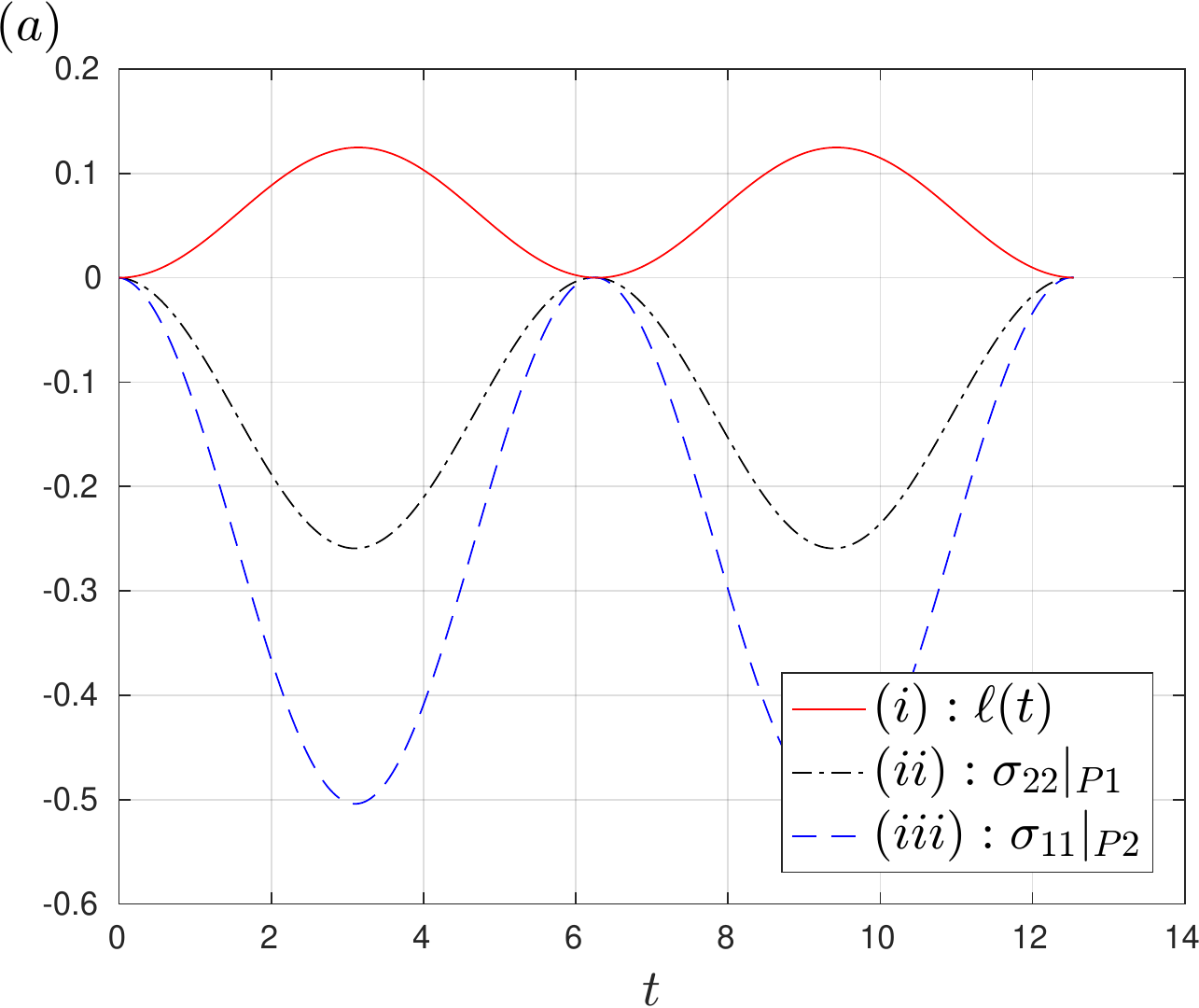}
\includegraphics[width=0.45\textwidth]{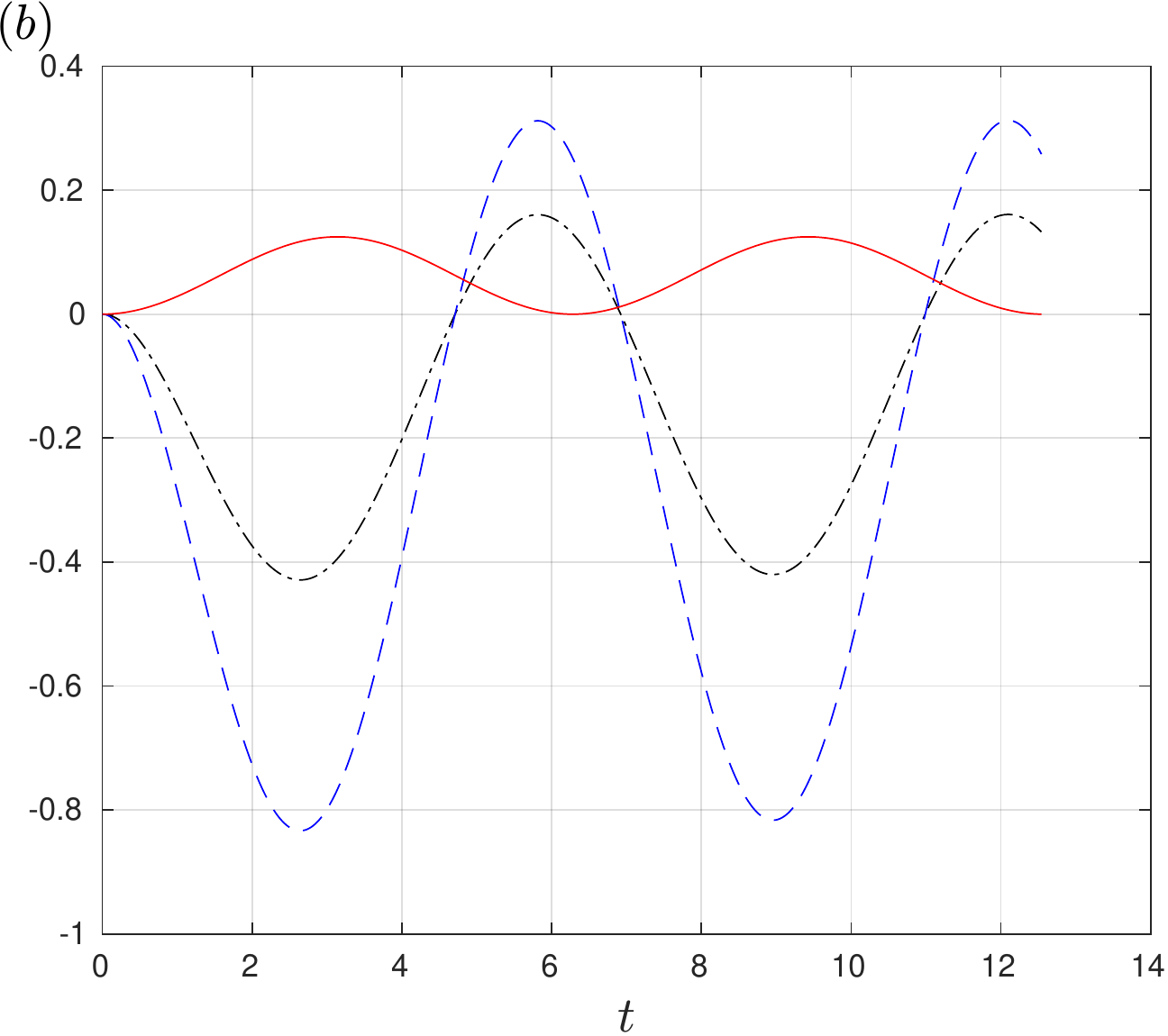}
\caption{The evolution of: (i) $\ell(t)$, the vertical displacement
  of the top of the unit cell $\Gamma_3^{\mathrm{mic}}$; (ii)
  $\sigma_{22}|_{P_1}$, the normal stress at the point $P_1$,
      and; (iii) $\sigma_{11}|_{P_2}$, the normal stress at the
      point $P_2$. The results shown are for anodes under slow (panel
      (a)) and fast (panel (b)) electrochemical cycling as discussed
      in the final paragraph of \S\ref{AM_swell}.}
\label{other_curves2}
\end{figure}

A similar pair of simulations (one slow cycling and one fast cycling)
were also carried out for an anode. The difference between the anode
and cathode is that, for an anode, the electrode particles are
initially delithiated (at their smallest) and their volume then
increases as the cell is discharged. Simulations for the anode were
carried out with $g(t) = (1-\cos(t))/10$. The evolution of the
stresses at $P_1$ and $P_2$ as well as $\ell(t)$ are shown in panels (a)
(for slow cycling with $K_\tau = G_\tau = 0.02$) and (b) (for fast
cycling with $K_\tau=G_\tau=1$) of Figure \ref{other_curves2}. For the
anode we find that the slow cycling protocol is unlikely to drive any
delamination at all --- mid-cycle the electrode particle is larger
than is was originally and therefore the surrounding binder is in
compression. Since the cycling timescale is much slower than the
relaxation timescale, very little creep occurs and so when the
particle returns to its original position the stresses on the surface
are negligibly small. Interestingly, and in contrast to the cathode,
the creep that is observed for the more aggressive C-rate (see panel
(b) of Figure \ref{other_curves2}) is likely to cause delamination
rather than prevent it. Here, when the particle returns to its
smallest (original) size the binder has crept such that it is now
under tension. Thus, fast cycling rates are likely to increase the
delamination damage in the anode. For similar reasons to those
discussed above, this creep has a greater effect on the right- and
left-hand sides than on the top and bottom of the particle surfaces.

\section{\label{discandconc}Discussion and conclusions}
We have developed a continuum mechanics model capable of predicting
the mechanical response of porous lithium-ion battery electrodes to
both (i) the swelling of the polymer binder on absorption of
electrolyte and; (ii) cyclic swelling/contraction of the electrode
particles as the battery is charged and discharged. The main goal of
this work has been to understand the causes of delamination of the
binder from the electrode particle surfaces as these devices are
assembled and used. We have shown that for typical electrodes, the
permeability is large enough that, on the time- and length-scales of
interest, pressure gradients within the electrolyte equilibrate
almost instantaneously, and do so without inducing any significant
stress within the surrounding porous skeleton. Crucially, this
decouples the fluid and solid components of the model, and reduces the
problem to one of predicting the mechanical state of the polymer
binder from a system of viscoelastic equations only. In order to
account for the complex geometry of a realistic electrode we employed
a multiscale approach to solving the decoupled model. First, the
governing equations were upscaled, in an ad-hoc manner, to give an
approximate homogeneous problem applicable on the electrode
lengthscale. The solution of the resulting macroscopic problem was
found to be almost entirely one-dimensional throughout the bulk of the
electrode --- a consequence of its slender geometry and its bonding to a
rigid current collector. This simple solution on the electrode
lengthscale was used to infer boundary conditions for the microscale
problem about an individual electrode particle. The solution of this
microscale problem was determined numerically using finite
elements techniques on a representative `unit cell'.

We demonstrated that the two different driving mechanisms, polymer
swelling and electrode particle volume changes, cause distinct modes
of deformation. Swelling of the polymer, caused by absorption of the
electrolyte, leads to large tensile stresses on the top and bottom
surfaces of the electrode particles which are likely to lead to
delamination from these surfaces. Cycling of the cathode, which is
constructed from electrode particles in their fully-lithiated
(largest) state, leads to tensile stresses along the surface of the
particle, but most strongly on its sides. This is expected to lead to
delamination occurring predominantly on the particle sides which
contrasts with the expected mode of delamination due to binder
swelling (along the top and bottom particle surfaces). In anodes, on
the other hand, in which the particles are initially delithiated
(smallest), cycling at slow rates, where the timescale for
(dis-)charge are very much longer than the viscoelastic relaxation, is
likely to cause little or no delamination. In high current
applications, however, it is possible to reach regimes where the creep
of the polymer becomes important and delamination can be induced in
the anode microstructure. These insights, combined with suitable
microscopy techniques, allow educated postmortems of real devices to
be carried out. If delamination is seen mainly at the upper and lower
surfaces of the electrode particles, then binder swelling is likely
the primary cause whereas if delamination has occurred at the lateral
edges it is more likely a result of cell cycling.

In cathodes delamination caused by binder swelling and cell cycling in
low current applications (where time scales for cycling are much
longer than those for viscoelastic relaxation) can be mitigated by:
(i) decreasing the volumetric changes associated with binder swelling
and electrode particle swelling/contraction and (ii) by decreasing the
long-term moduli of the polymer --- the other material properties of
the binder have little or not effect. For high current applications,
the delamination induced in the anode can also be mitigated by using
the tactics discussed above, or by decreasing either the short-term
moduli or further decreasing the characteristic relaxation time of the
polymer.

Returning for a moment to the images shown at the start of this study
--- see Figure \ref{example_slices} --- we can now rationalize the
different stages of morphological deterioration. In panel (a), prior
to soaking in electrolyte, we see that the binder is well attached to
the electrode particle surfaces. In panel (b), after electrolyte
immersion but before cycling, we see significant delamination from the
upper and lower parts of the electrode particle surfaces only. A
simple visual inspection of panel (c) also reveals large regions of
delamination running left to right. However, a more detailed analysis
of these same images was carried out in \cite{nester}, and there it
was found that a measurable increase in the amount of delamination
(around 5\% in terms of electrode particle surface area) had occurred
as a direct result of the electrochemical cycling stages. All three of
these observations are consistent with the results of our model. The
cycling stages caused significantly less damage than the swelling of
the polymer for the electrodes shown in Figure \ref{example_slices}.
This is likely due to the NMC active material which only changes its
volume a small amount ($\sim$2-4\%) on (de-)lithiation. However, we do
emphasize that for other chemistries, particularly newer silicon-based
materials, which can exhibit extremely large volumetric expansions,
the mechanical damage caused by active material swelling could be far
more significant.

In \cite{powfollow} the framework developed here has already been used to explore the effects of alterations in particle shape and differing polymer rheology. Perhaps the most significant result in \cite{powfollow} was the demonstration that the size of the deterimental tensile normal stresses on the top and bottom surfaces of the particles caused by binder swelling can be reduced by using non-circular particles, e.g. ellipses, and aligning the longer edges of the particles in planes parallel with the CC. On the one hand this distributes the damaging tensile stress over a larger area, thereby reducing its size, but a larger area is then exposed to this relatively smaller stress. Before comments can be made on optimal particle shape, it is therefore crucial to understand the size of the threshold stress required to cause delamination. Other interesting open questions that remain to be addressed include: (i) formalizing the upscaling arguments used in \S\ref{the_planet}; (ii) using the model and methods developed here to discern the effects of altering the packing of the electrode particles. Results on both these fronts will be reported in the near future.

\section*{Acknowledgements}
Thanks to S. Krachkovskiy and G. Goward for constructing and cycling the cells, and to H. Liu and G. Botton for providing the microscopy images used in this study. Thanks also to K. Harris, X. Huang, J. H. Kim, I. Halalay, C. P. Please and I. Danaila for stimulating discussions regarding this work. The electron microscopy data included in this paper was acquired at the
Canadian Centre for Electron Microscopy, a national facility supported
by the Natural Sciences and Engineering Research Council of Canada
(NSERC) and McMaster University. Funding for this research was provided
by Automotive Partnership Canada (APC) and General Motors of Canada.

\end{document}